\documentclass[conference,compsoc]{IEEEtran}
\ifCLASSOPTIONcompsoc
  \usepackage[nocompress]{cite}
\else
  \usepackage{cite}
\fi
\ifCLASSINFOpdf
\else
\fi
\usepackage{amsmath}
\usepackage{amsthm}
\usepackage{amsmath,amssymb}
\usepackage{graphicx}
\usepackage{booktabs}
\usepackage{pifont}
\usepackage{tabularx}
\usepackage[table]{xcolor}
\usepackage{multirow}
\usepackage[ruled,linesnumbered]{algorithm2e}
\usepackage[table]{xcolor}
\usepackage{enumitem}
\usepackage{xurl}
\usepackage{hyperref}
\usepackage{marvosym}

\usepackage[most]{tcolorbox}
\usepackage{listings}
\usepackage{xcolor}

\newtcblisting{casebox}[1]{
    enhanced,
    listing only,
    listing engine=listings,
    colback=gray!5,
    colframe=black!55,
    colbacktitle=black!75,
    coltitle=white,
    fonttitle=\bfseries\footnotesize,
    title={#1},
    boxrule=0.5pt,
    arc=1mm,
    boxsep=0pt,
    left=1.5mm,
    right=1.5mm,
    top=1mm,
    bottom=1mm,
    height=2cm,
    toptitle=0.8mm,
    bottomtitle=0.8mm,
    before skip=0pt,
    after skip=0pt,
    listing options={
        language=Python,
        basicstyle=\ttfamily\scriptsize,
        keywordstyle=\color{black},
        stringstyle=\color{black},
        commentstyle=\color{gray},
        columns=fullflexible,
        keepspaces=true,
        breaklines=false,
        showstringspaces=false,
        aboveskip=0pt,
        belowskip=0pt
    }
}

\makeatletter
\@ifundefined{theorem}
  {\newtheorem{theorem}{Theorem}}{}
\@ifundefined{lemma}
  {}{}
\@ifundefined{proposition}
  {\newtheorem{proposition}[theorem]{Proposition}}{}
\@ifundefined{assumption}
  {}{}
\makeatother

\SetKwComment{Comment}{$\triangleright$\ }{}
\DontPrintSemicolon

\begin{document}
\title{``More Than You Expect!'': Leakage-Abuse Attacks on Encrypted Vector Search}

% \author{\IEEEauthorblockN{Michael Shell}
% \IEEEauthorblockA{School of Electrical and\\Computer Engineering\\
% Georgia Institute of Technology\\
% Atlanta, Georgia 30332--0250\\
% Email: http://www.michaelshell.org/contact.html}
% \and
% \IEEEauthorblockN{Homer Simpson}
% \IEEEauthorblockA{Twentieth Century Fox\\
% Email: homer@thesimpsons.com}
% \and
% \IEEEauthorblockN{James Kirk\\ and Montgomery Scott}
% \IEEEauthorblockA{Starfleet Academy\\
% San Francisco, California 96678-2391\\
% Telephone: (800) 555--1212\\

% \author{\IEEEauthorblockN{Anonymous}
% \IEEEauthorblockA{Anonymous Institution}}

\author{
    Yang Zhang\textsuperscript{$\spadesuit \ *$},
    Ruiqi He\textsuperscript{$\spadesuit \ *$}, 
    Xiaoqian Sun\textsuperscript{$\spadesuit$}, 
    Siyi Lv\textsuperscript{$\spadesuit$\ \Letter}, 
    Zheli Liu\textsuperscript{$\spadesuit$}, 
    \\[6pt]
    \textsuperscript{$\spadesuit$}\textit{CS\&CCS, DISSec, AAIS, Nankai University,} \\
    \textit{\{yangz, heruiqi, sunxiaoqian\}@mail.nankai.edu.cn, \{lvsiyi, liuzheli\}@nankai.edu.cn}

    \thanks{\textsuperscript{*} \textup{Equal Contribution.}}
    \thanks{\textsuperscript{\Letter} \textup{Corresponding Author.}}
}


% compsoc conferences compared to traditional conferences), use this
%Telephone: (800) 555--1212, Fax: (888) 555--1212}


\maketitle


\begin{abstract}

Vector databases support semantic search and retrieval-augmented generation, but the embeddings, queries, and record labels they handle may contain sensitive information. Encrypted vector search aims to protect this information. However, some encrypted vector search schemes fail to hide which records are retrieved together across queries. We identify \emph{semantic neighborhood leakage}, showing that co-retrieval relationships derived from access patterns can reveal the semantic categories of encrypted records and queries without exposing their underlying contents.

We present SAGE (Semantic Access-pattern Graph Exploitation), a passive and persistent leakage-abuse attack that infers record labels and query categories from access patterns by exploiting semantic neighborhood leakage. SAGE constructs a weighted co-retrieval graph and develops two independent attack variants under distinct auxiliary-knowledge assumptions. SAGE-Dist matches anonymous spectral groups to semantic categories using estimated class proportions and refines record labels via optimal transport, without requiring known record labels. SAGE-Known combines graph-based label propagation with anchor voting over spectral groups, using a small set of known record labels.

We evaluate SAGE on four datasets spanning text and images. With only 1\% known record labels, SAGE-Known achieves up to 84.3\% mean record-label accuracy and consistently outperforms the strongest corresponding baselines across all four datasets. Using only estimated class proportions, SAGE-Dist achieves 66.9\% record-label accuracy and 68.1\% query-category accuracy on AG News, exceeding the strongest corresponding baselines by 21.8 and 41.9 percentage points, respectively. However, its effectiveness varies across datasets. Experiments with a real-world DCPE implementation (Fuchsbauer et al., SCN'22) further show that semantic leakage persists under encrypted approximate retrieval. Our findings highlight the privacy risks posed by semantic neighborhood leakage, even when vector contents remain protected.

\end{abstract}


% \fi
\IEEEpeerreviewmaketitle

\section{Introduction}
\label{sec:intro}

Vector databases have become an important component of applications that search and organize unstructured information~\cite{pan2023,taipalus2024vector,10.1145/3626246.3654691,DBLP:journals/corr/abs-2310-11703}. They support retrieval-augmented generation (RAG)~\cite{DBLP:conf/nips/LewisPPPKGKLYR020,DBLP:journals/corr/abs-2312-10997,DBLP:journals/csur/HuangH26}, semantic search~\cite{karpukhin2020dense,reimers2019sentencebertsentenceembeddingsusing}, recommendation~\cite{huang2020embedding,he2023que2engage}, and data annotation~\cite{gonzalez-gutierrez-quattoni-2024-leveraging,10.1145/3711542.3711570,eldin2026semanticlabelingthirdpartycybersecurity}. In the setting studied here, each database record contains four fields: its original content (e.g., text and images), an embedding derived from that content, a semantic label and a stable record identifier (ID). Rather than matching labels, the system retrieves the top-k nearest records based on embedding similarity. Since semantically related records often lie close together in the embedding space and share category labels~\cite{gonzalez-gutierrez-quattoni-2024-leveraging,10.1145/3711542.3711570}, the resulting retrieval neighborhoods can reflect the underlying semantic structure of the database.

\begin{figure}[t]
    \centering
    \includegraphics[width=\columnwidth]{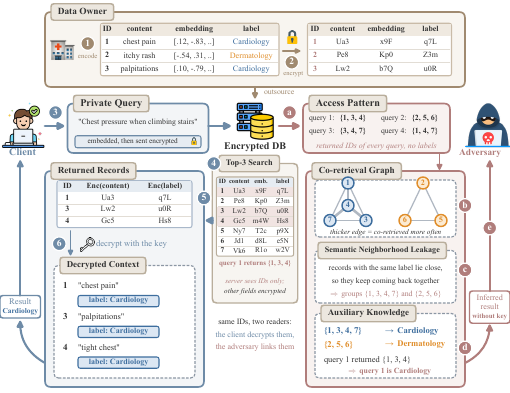}
    \caption{Semantic neighborhood leakage in encrypted
vector search. While record contents, embeddings,
queries, and semantic labels remain protected,
the server observes record identifiers
returned by legitimate queries. As illustrated,
Queries 1 and 3 both retrieve records 3 and 4, exposing co-retrieval relationships.
By aggregating such relationships across queries,
an adversary can exploit semantic neighborhood
leakage to infer record labels and query
categories with limited auxiliary knowledge.}
    \label{fig:teaser}
\end{figure}

However, exposing database and query embeddings in plaintext can pose significant privacy risks. Embedding inversion attacks can reconstruct original text from accessible embeddings, revealing sensitive information~\cite{morris2023text,li2023sentence,wan2024information}. Moreover, plaintext semantic labels directly disclose sensitive categories. These risks motivate encrypted similarity-search techniques for protecting database and query embeddings, together with separate encryption of sensitive record content and labels. Figure~\ref{fig:teaser} illustrates such a deployment
using clinical records whose text, embeddings, and
diagnostic labels are protected before outsourcing,
while their stable record identifiers remain
linkable across queries.

Encryption does not necessarily conceal how records are accessed. We refer to
the set of returned record identifiers for each
query as its \emph{access pattern}. Some encrypted
vector-search schemes hide retrieval accesses
through PIR or oblivious computation~\cite{zhou2025pacmann,li2025panther,asi2024wally,
boldyreva2021privacy}, but these approaches
introduce additional computation, communication,
or deployment costs. Other schemes prioritize
efficient retrieval and expose record
identifiers~\cite{liu2025privacy,wang2026efficient,
wong2009secure,kuzu2012efficient}. Such exposure also occurs in practical systems,
as demonstrated by our experiments with Quiver~\cite{soua2026quiver},
a vector search system supporting
distance-comparison-preserving encryption (DCPE)~\cite{fuchsbauer2022approximate}. Appendix~\ref{app:identifier_exposure}
further examines identifier exposure in practical
vector search systems. Following Hoover
et al.~\cite{hoover2024leakage} and Sun
et al.~\cite{sun2026can}, we consider a
\emph{passive and persistent} server-side adversary
who observes returned record identifiers but
cannot access plaintext record contents,
embeddings, queries, or semantic labels.
Although these identifiers need not carry
semantic meaning themselves, repeated
observations reveal overlaps across query results.

Such overlaps can reveal semantic relationships
among encrypted records. We say that two records
are \emph{co-retrieved} when they appear in the
result of the same query. Since top-$k$ retrieval
reflects proximity in an embedding space,
repeated co-retrieval can expose latent semantic
neighborhoods. In Figure~\ref{fig:teaser},
the returned sets $\{1,3,4\}$ and $\{3,4,7\}$
share records $3$ and $4$. By aggregating
such relationships across queries, an adversary
can recover anonymous groups of semantically
related records. With limited auxiliary knowledge,
such as estimated class proportions or a few
known labels, these groups can further reveal
\emph{record labels} and \emph{query categories}.
We refer to this category-level exposure as
\textbf{semantic neighborhood leakage}.
In the clinical-record example, an adversary
may associate a recovered group with cardiology,
thereby exposing the diagnostic categories of
protected records and the medical topics of
private queries. This inference does not require
reconstructing record contents or embeddings,
nor does it require every co-retrieved record
to share the same label.

The possibility of inferring hidden semantic
information raises the question of whether existing
leakage-abuse attacks can exploit such leakage.
Prior work has demonstrated some inference
attacks against encrypted search~\cite{islam2012access,cash2015leakage,
naveed2015inference}.
These attacks exploit numerical ordering and
statistical relationships to infer protected
information. However, existing techniques
do not directly address semantic-category
inference from high-dimensional top-$k$
similarity search.

\subsection{Limitations of Existing Attacks}
\label{sec:limits}

\noindent\textbf{Order-based inference.}
Existing attacks on encrypted numerical databases exploit
ordering constraints exposed by range and nearest-neighbor
queries~\cite{kellaris2016generic,kornaropoulos2019data,
kornaropoulos2020state}. In one-dimensional $k$-NN search,
returned records form consecutive intervals in a common
sorted order, allowing adversaries to reconstruct their
relative positions from overlapping access patterns.
Prior work has also extended reconstruction to
multidimensional range queries, including attacks that
exploit co-occurrence relationships and graph drawing
to recover geometric structure~\cite{falzon2020full,markatou2021reconstructing,
markatou2023attacks,markatou2024reconstructing}.
\emph{However, high-dimensional top-$k$ similarity search
does not expose a global record ordering or the
axis-aligned constraints of range queries.} Its returned
sets instead reflect local neighborhoods in an embedding
space, making existing reconstruction techniques
difficult to apply directly. Moreover, recovering
geometric relationships does not identify the semantic
categories of records or queries.

\noindent\textbf{Statistical matching attacks.}
Attacks on encrypted keyword and equality search
exploit frequencies and co-occurrence statistics to
match observable leakage with candidate plaintext
values~\cite{islam2012access,cash2015leakage,
oya2022ihop,nie2024query,sun2026can}.
These techniques typically rely on corresponding
auxiliary statistics for candidate keywords or values.
In contrast, vector similarity search retrieves records
according to embedding proximity rather than exact
predicate matches. Queries from the same semantic
category may retrieve different records, while queries
from different categories may return overlapping sets.
\emph{Consequently, co-retrieval relationships provide noisy
evidence of semantic similarity rather than a direct
correspondence to plaintext keyword or value statistics.}
Furthermore, even if anonymous record groups are
recovered, access patterns alone cannot identify their
semantic categories, since permuting the hidden category
names leaves the observed access patterns unchanged.

\subsection{Our Contributions}
\label{sec:contrib}

We present SAGE (Semantic Access-pattern Graph
Exploitation), a passive and persistent leakage-abuse
attack that exploits semantic neighborhood leakage
to infer record labels and query categories from
access patterns in encrypted vector search.
Our contributions are as follows.

\begin{itemize}[leftmargin=*,align=parleft]

    \item \textbf{Characterizing semantic neighborhood leakage.}
    We investigate how repeated top-$k$ retrievals
    expose semantic relationships among encrypted
    records through observable identifiers.
    We construct a weighted co-retrieval graph
    to capture these relationships and examine
    the semantic structure revealed by spectral
    grouping. We distinguish the recovery of
    anonymous record groups from the identification
    of their semantic categories.

    \item \textbf{Two attacks with distinct auxiliary knowledge.}
    We develop two independent inference attacks.
    SAGE-Dist uses estimated class proportions
    to associate anonymous groups with semantic
    categories, followed by optimal-transport
    refinement of record labels.
    SAGE-Known combines graph-based label
    propagation, class-mass normalization, and
    anchor voting over spectral groups using
    a small set of known record labels.
    Both attacks infer query categories from
    their predicted record labels.

    \item \textbf{Evaluation and privacy implications.}
    We evaluate SAGE on four datasets spanning text
    and images under varying observation workloads
    and auxiliary-knowledge settings. With 1\% known
    record labels, SAGE-Known achieves mean record-label
    accuracies of 84.3\%, 64.9\%, 47.0\%, and 27.8\%
    across the four datasets. Using only estimated
    class proportions, SAGE-Dist achieves 66.9\%
    on AG News and 36.1\% on Yahoo Answers, although
    its effectiveness varies substantially across
    datasets and runs. Experiments with Quiver DCPE
    further demonstrate that semantic inference
    can remain effective under encrypted approximate
    retrieval, highlighting the privacy risks of
    semantic neighborhood leakage.

\end{itemize}

\section{Related Work}
\label{sec:rw}

\noindent\textbf{Leakage-abuse attacks on encrypted search.}
Prior work has demonstrated that search,
access, and volume patterns can reveal
sensitive information even when database contents
and queries are encrypted~\cite{islam2012access,cash2015leakage,
naveed2015inference,oya2021hiding}.
Attacks against encrypted numerical databases exploit
ordering and response overlaps to
reconstruct values and geometric relationships~\cite{kellaris2016generic,cryptoeprint:2017/701,
grubbs2018pump,markatou2019full,falzon2020full,
markatou2021reconstructing,markatou2023attacks,
markatou2024reconstructing}.
Kornaropoulos et al.~\cite{kornaropoulos2019data,
kornaropoulos2020state} further investigate
reconstruction attacks against one-dimensional
nearest-neighbor search.
In encrypted keyword search, IHOP~\cite{oya2022ihop}
and Jigsaw~\cite{nie2024query} exploit statistical
relationships and auxiliary knowledge to recover
protected queries.
Hoover et al.~\cite{hoover2024leakage} investigate leakage-abuse attacks against structured
encryption for SQL, while
Sun et al.~\cite{sun2026can} develop incremental
inference attacks that exploit correlations and
joint distributions across encrypted SQL attributes.
Other attacks similarly leverage frequency and
co-occurrence statistics~\cite{damie2021highly}.
While these attacks exploit ordering constraints
or statistical relationships in structured search,
SAGE investigates semantic-category
inference from co-retrieval relationships in
high-dimensional top-$k$ similarity search.
The limitations of existing attacks in this setting
are discussed in Section~\ref{sec:limits}.

\noindent\textbf{Privacy-preserving vector search.}
Some private nearest-neighbor schemes employ oblivious
data structures, private information retrieval, or
other privacy-preserving mechanisms to protect
retrieval patterns~\cite{boldyreva2021privacy,zhou2025pacmann,
li2025panther,asi2024wally}.
Other approaches support efficient server-side search
over protected vectors.
Early schemes employ scalar-product-preserving
encryption or locality-sensitive hashing for
encrypted nearest-neighbor and similarity search~\cite{wong2009secure,kuzu2012efficient}.
More recently, Liu et al.~\cite{liu2025privacy}
combine encrypted distance comparisons with
HNSW-based approximate search, while
Wang et al.~\cite{wang2026efficient} extend this
approach to range-filtered retrieval and explicitly
allow final result identifiers to be exposed.
SAGE targets deployments where the server can
observe and link the selected record identifiers
across queries, without requiring plaintext
embeddings, distance values, or index traversal traces.

\noindent\textbf{Privacy attacks on embeddings and retrieval systems.}
When plaintext embeddings are accessible, embedding inversion
attacks can reconstruct information about their
source inputs~\cite{morris2023text,li2023sentence,
wan2024information}.
Other studies investigate membership inference
against retrieval-augmented generation systems
through interactions or externally observable outputs~\cite{anderson2024my,li2025generating,
liu2025mask,naseh2025riddle}.
In contrast, SAGE requires neither plaintext
embeddings nor retrieved content or generated
answers. It exploits co-retrieval relationships
derived from observable access patterns to infer
the semantic labels of database records and the
categories of user queries.

\section{Problem Statement}
\label{sec:problem}

\subsection{Encrypted Vector Search}
\label{sec:encrypted-search}

We consider an outsourced encrypted vector-search
service involving a data owner, an untrusted server,
and authorized clients, as shown in
Figure~\ref{fig:teaser}.
The database contains $N$ records
$\mathcal{R}=\{r_1,\ldots,r_N\}$, where
\begin{equation}
    r_j=(\mathrm{id}_j,b_j,\lambda_j,x_j),
    \qquad j\in\{1,\ldots,N\}.
    \label{eq:record}
\end{equation}
Here, $\mathrm{id}_j$ is the unique record
identifier, $b_j\in\mathbb{R}^{d}$ is its embedding
derived from the original content $x_j$, and
$\lambda_j\in\mathcal{C}$ is its semantic label.
The label space $\mathcal{C}=\{0,\ldots,C-1\}$
contains $C$ mutually exclusive categories.
The index $j$ enumerates records, whereas
$\mathrm{id}_j$ denotes their actual identifiers.
Let $\mathcal{B}=\{b_1,\ldots,b_N\}$ denote
the database embeddings.

The encrypted retrieval process consists of
six steps.
\ding{182}~\textit{Local encoding.}
The owner encodes record contents into embeddings
and assigns record identifiers and labels.
\ding{183}~\textit{Encrypted outsourcing.}
The owner outsources protected embeddings and
their associated identifiers.
Randomized encryption may produce different
ciphertexts for identical plaintexts without
hiding stable record identifiers.
\ding{184}~\textit{Query encryption.}
For query $i$, the client generates an embedding
$a_i\in\mathbb{R}^{d}$ and submits its protected
representation. Let $\ell_i\in\mathcal{C}$
denote the query's semantic category.
\ding{185}~\textit{Encrypted similarity search.}
The server performs protected top-$k$ retrieval.
Let $\operatorname{TopK}(a_i,\mathcal{B})$
denote the indices of the $k$ records selected
for query $a_i$. The returned identifier set is
\begin{equation}
    I_i=\{\mathrm{id}_j\mid
    j\in\operatorname{TopK}(a_i,\mathcal{B})\},
    \qquad |I_i|=k.
    \label{eq:topk}
\end{equation}
Each element of $I_i$ is a record identifier.
\ding{186}~\textit{Result retrieval.}
The server returns the selected identifiers
and any requested protected record contents.
\ding{187}~\textit{Local decryption.}
The client processes the retrieved results
and decrypts any returned ciphertexts.

The selected identifiers may remain visible
to the server during
Steps~\ding{185} and~\ding{186}.

\subsection{Threat Model and Leakage}
\label{sec:threat}

\noindent\textbf{Adversary.}
Following prior leakage-abuse attacks~\cite{hoover2024leakage},
we consider a \emph{passive and persistent}
server-side adversary that observes legitimate
retrieval executions without interfering with them.
The adversary has no decryption keys and cannot
access plaintext record contents, database
embeddings, semantic labels, or query contents and embeddings,
except for independently obtained auxiliary knowledge.
It neither injects queries nor modifies the
database or retrieval results.

\noindent\textbf{Access-pattern leakage.}
Suppose the server processes $M$ legitimate queries.
For each query $i\in\{1,\ldots,M\}$, the adversary
observes the unordered set $I_i$ of selected record
identifiers defined in Eq.~\eqref{eq:topk}.
The accumulated \emph{access-pattern leakage} is
\begin{equation}
    \mathcal{L}_{\mathrm{AP}}=(I_1,I_2,\ldots,I_M).
    \label{eq:access-leakage}
\end{equation}
The adversary only needs to recognize repeated
identifiers across queries. Thus, opaque or
encrypted identifiers do not prevent this leakage
if they remain linkable. SAGE uses neither
similarity scores nor retrieval rankings. Our attack applies to deployments exposing
cross-query-linkable result identifiers.

\noindent\textbf{Semantic neighborhood leakage.}
Similarity search retrieves records based on
embedding proximity. As semantically related
records often occupy nearby regions of the
embedding space, they may repeatedly appear
together in query results.
For each record $r_j$, we define its
\emph{retriever set} as the indices of
observed queries that return its identifier:
\begin{equation}
    Q_j=\{i\in[M]\mid \mathrm{id}_j\in I_i\}.
    \label{eq:retriever-set}
\end{equation}
Here, $|Q_j|$ denotes the retrieval frequency
of record $r_j$, while $|Q_j\cap Q_\ell|$
counts the queries that retrieve both records
$r_j$ and $r_\ell$. These observable co-retrieval relationships
can expose the database's latent semantic
structure, giving rise to semantic neighborhood
leakage. The exposed structure does not
necessarily identify semantic categories
without auxiliary knowledge.

\subsection{Adversarial Goal}
\label{sec:adversarial-goal}

\noindent\textbf{Attack objective.}
The adversary aims to infer the record labels
of observed database records and the categories
of legitimate queries.
Let $V=\bigcup_{i=1}^{M}I_i$ denote the set
of distinct record identifiers observed
during retrieval.
Given the access-pattern leakage
$\mathcal{L}_{\mathrm{AP}}$ and auxiliary
knowledge $\mathcal{P}$, SAGE performs
\begin{equation}
    \mathrm{SAGE}:
    (\mathcal{L}_{\mathrm{AP}},\mathcal{P})
    \longrightarrow
    (\hat{\boldsymbol{\lambda}},
     \hat{\boldsymbol{\ell}}),
    \label{eq:attack-goal}
\end{equation}
where $\hat{\boldsymbol{\lambda}}$ contains
the predicted labels of records whose
identifiers belong to $V$, and
$\hat{\boldsymbol{\ell}}$ contains the
predicted categories of the $M$ observed
queries. Records outside $V$ are not
considered in the attack.

Throughout this paper, \emph{record label}
refers to $\lambda_j$, and \emph{query category}
refers to $\ell_i$. Both take values in the
same set of semantic categories $\mathcal{C}$.

\noindent\textbf{Auxiliary knowledge.}
We consider two settings.
In \emph{SAGE-Dist}, the adversary knows
estimated class proportions
$\mathcal{P}=\hat{\boldsymbol{\pi}}$,
where
$\hat{\boldsymbol{\pi}}
=(\hat{\pi}_c)_{c\in\mathcal{C}}$
specifies the estimated database proportion
of each named semantic category.
These estimates may be derived from an
independent labeled dataset or publicly
available statistics. For example, AHRQ's
HCUP Fast Stats provides diagnosis-frequency
statistics that could inform class-proportion
estimates for clinical databases with
compatible diagnostic categories~\cite{ahrq_hcup}.
No known labels linked to individual target
records are required.
In \emph{SAGE-Known}, the adversary instead
possesses a small set of known identifier-label
pairs,
\begin{equation}
    \mathcal{P}=\Psi
    \subseteq
    \{(\mathrm{id}_j,\lambda_j)
    \mid 1\leq j\leq N\},
    \label{eq:known-labels}
\end{equation}
where each pair associates a record label
with its corresponding server-visible record
identifier.
For example, arXiv provides publicly accessible
paper identifiers and subject categories
through its metadata API~\cite{arxiv_api}.
If an encrypted paper-search database retains
these identifiers, externally obtained metadata
can provide known identifier-label pairs.
We assume these associations are obtained
independently of the attack, without query
injection or database modification.
Only pairs in $\Psi$ whose identifiers belong
to $V$ can serve as labeled anchors in the
observed co-retrieval graph.
SAGE-Known does not require class-proportion
estimates.

Both settings assume a known number $C$
of named semantic categories, as may arise
from public taxonomies or standardized
classification systems.
We consider a fixed, single-label
categorization at a chosen granularity.
Not every category must appear among the
observed records.
Unknown category counts and multilabel
categorization fall outside this formulation.

\section{Technique Overview}
\label{sec:overview}

SAGE exploits semantic neighborhood leakage
to infer record labels and query categories
from observable access patterns. It first
constructs a co-retrieval graph that captures
relationships among records retrieved together
across queries. These relationships reveal
structural groups without identifying the
semantic categories of their records. SAGE then
uses one of two forms of auxiliary knowledge
to infer record labels from the observed
structure. Finally, both attack variants
infer query categories from their predicted
record labels.

\subsection{Co-retrieval Graph Framework}
\label{sec:overview-graph}

\begin{figure*}[!t]
    \centering
    \includegraphics[width=\textwidth]{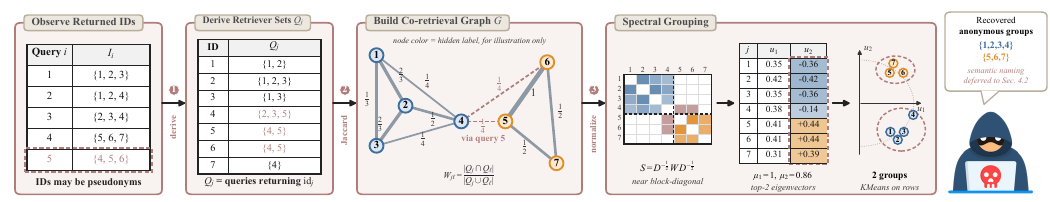}
    \caption{From access patterns to anonymous record
    groups. The adversary derives retriever sets
    $Q_j$ from the observed access sets $I_i$
    and constructs a weighted co-retrieval graph.
    Spectral grouping identifies related records
    without revealing their record labels.
    The spectral visualization is schematic.}
    \label{fig:co-retrieval-overview}
\end{figure*}

\noindent\textbf{Key observation.}
An individual returned identifier does not
directly reveal its record label.
Across queries, however, records belonging
to the same semantic category may be retrieved
together more frequently than records
from different classes.
Their co-retrieval relationships can therefore
expose semantic structure even when contents, embeddings and labels remain encrypted.

\noindent\textbf{Graph construction.}
SAGE represents the observed records as a weighted,
undirected graph $G=(V,E,W)$, where $V$ is the
set of observed record identifiers defined in
Section~\ref{sec:adversarial-goal}.
Two distinct records are connected if they
occur together in at least one observed
query result. SAGE assigns edge weights
using the Jaccard similarity of their
retriever sets:
\begin{equation}
    W_{j\ell}
    =\frac{|Q_j\cap Q_\ell|}
           {|Q_j\cup Q_\ell|},
    \qquad j\neq\ell,
    \label{eq:overview-jaccard}
\end{equation}
where $\mathrm{id}_j,\mathrm{id}_\ell\in V$
and $Q_j$ is defined in
Section~\ref{sec:threat}.
We set $W_{jj}=0$.
The matrix $W$ is indexed by a fixed
ordering of the observed identifiers.

\noindent\textbf{Illustrative example.}
As shown in Figure~\ref{fig:co-retrieval-overview},
consider seven encrypted medical records
and five observed queries with $k=3$.
For simplicity, the record identifiers
are denoted by $1,\ldots,7$.
The returned identifier sets are
$I_1=\{1,2,3\}$,
$I_2=\{1,2,4\}$,
$I_3=\{2,3,4\}$,
$I_4=\{5,6,7\}$, and
$I_5=\{4,5,6\}$.
Records 1 and 2 are retrieved together
in two queries, yielding $W_{12}=2/3$.
Records 5 and 6 always occur together,
yielding $W_{56}=1$.
By contrast, records 4 and 5 share
only one query, yielding $W_{45}=1/4$.
These relationships form two relatively
cohesive groups, $\{1,2,3,4\}$ and
$\{5,6,7\}$, connected by weaker
cross-group edges.

Suppose the first group predominantly
contains cardiology records, while
the second predominantly contains
dermatology records.
The adversary can observe the two
groups but cannot determine their
predominant semantic categories from
the identifiers alone.

\noindent\textbf{Anonymous grouping.}
SAGE applies spectral analysis to the
normalized co-retrieval graph to obtain
structural representations of observed
records. Spectral grouping identifies
anonymous groups whose members exhibit
similar co-retrieval relationships.
These groups need not correspond exactly
to semantic categories, and their members'
record labels remain unknown.

The central challenge is therefore
to infer record labels from anonymous
graph structure using auxiliary knowledge.
SAGE addresses this challenge through
two distinct inference strategies.
Section~\ref{sec:graph-construction}
presents the graph construction
and spectral procedures.

\subsection{SAGE Attack Overview}
\label{sec:overview-attack}

\begin{figure*}[!t]
    \centering
    \includegraphics[width=\textwidth]{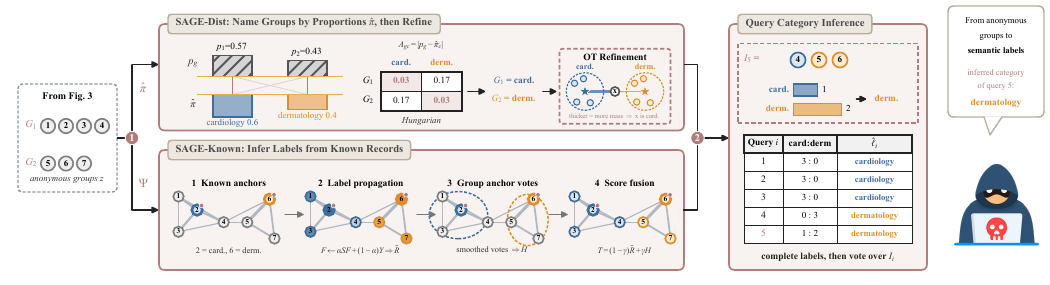}
    \caption{Overview of SAGE. SAGE-Dist assigns
    semantic labels using estimated class proportions.
    SAGE-Known combines graph-based label propagation
    with group-level anchor voting. Both variants
    infer query categories from predicted record labels.}
    \label{fig:sage-overview}
\end{figure*}

The co-retrieval graph reveals structural relationships
among records but does not directly reveal their
semantic categories. SAGE develops two attack variants
that exploit this structure using distinct forms of
auxiliary knowledge, as illustrated in
Figure~\ref{fig:sage-overview}.

\noindent\textbf{SAGE-Dist.}
SAGE-Dist assumes estimated class proportions
$\hat{\boldsymbol{\pi}}$, without knowing any
individual record labels. It first partitions
the co-retrieval graph into $C$ anonymous groups
using spectral embedding and KMeans. It then
matches the relative group sizes to the estimated
class proportions using the Hungarian algorithm
to assign semantic classes to the groups.
To account for imperfect grouping, SAGE-Dist
further refines individual record-label predictions
through Sinkhorn-based optimal transport, which
constrains the soft assignment masses according
to the estimated class proportions.

For example, consider the two anonymous groups
in Figure~\ref{fig:co-retrieval-overview}.
If cardiology is estimated to be more prevalent
than dermatology, the larger group can be
associated with cardiology and the smaller
with dermatology.

\noindent\textbf{SAGE-Known.}
SAGE-Known instead assumes a small set of known
identifier-label pairs $\Psi$. Known records
present in the observed graph serve as anchors.
SAGE-Known propagates their labels through
weighted co-retrieval connections while keeping
anchor labels fixed. The propagated scores are
normalized by class-wise mass to reduce class
imbalance. In parallel, spectral grouping provides
anonymous groups in which known anchors contribute
smoothed class votes. The two sources of evidence
are combined to infer the labels of other records.

For example, if record 2 is a known cardiology
record and record 6 is a known dermatology
record, their labels provide evidence for
labeling neighboring records through graph
connections and group memberships.
Section~\ref{sec:attack} details both procedures.

\noindent\textbf{Query category inference.}
After completing record-label predictions,
both variants infer each observed query's
category by majority voting over the predicted
labels of its retrieved records.
For instance, query 5 retrieves
$\{4,5,6\}$. If record 4 is predicted as
cardiology and records 5 and 6 as dermatology,
the inferred query category is dermatology.
This inference estimates the query's semantic
category.
The completion and voting procedures are
specified in Section~\ref{sec:attack}.

\section{Co-retrieval Graph Construction}
\label{sec:graph-construction}

Section~\ref{sec:overview-graph} illustrates how
repeated retrievals expose relationships among
otherwise opaque records. We now formalize the
construction of the co-retrieval graph and the
recovery of anonymous record groups.

\subsection{Constructing the Co-retrieval Graph}
\label{sec:graph-building}

\noindent\textbf{Query--record incidence.}
Given the observed access patterns
$\mathcal{L}_{\mathrm{AP}}=(I_1,\ldots,I_M)$,
SAGE constructs a binary incidence matrix
$X\in\{0,1\}^{M\times N}$, where
\begin{equation}
    X_{ij}
    =\mathbf{1}[\mathrm{id}_j\in I_i].
    \label{eq:incidence}
\end{equation}
Each row corresponds to an observed query,
and each column indicates whether a record
was retrieved. The number of times record $r_j$ is retrieved
is given by $|Q_j|=\sum_{i=1}^{M}X_{ij}$.

Recall that $V=\bigcup_{i=1}^{M}I_i$ denotes
the set of distinct server-visible record
identifiers observed across queries.
Let $N_{\mathrm{obs}}=|V|$ and
$X_V\in\{0,1\}^{M\times N_{\mathrm{obs}}}$
denote the submatrix containing the columns
of records whose identifiers belong to $V$,
arranged in a fixed order.
All subsequent graph matrices follow this
ordering.

\noindent\textbf{Co-retrieval weights.}
The matrix $O=X_V^{\mathsf T}X_V$
captures pairwise co-retrieval counts.
For two distinct observed records
$r_j$ and $r_\ell$, the corresponding
entry satisfies
$O_{j\ell}=|Q_j\cap Q_\ell|$.

SAGE constructs a weighted, undirected
graph $G=(V,E,W)$, where $W$ is the
Jaccard-weighted adjacency matrix:
\begin{equation}
W_{j\ell}=
\begin{cases}
\displaystyle
\frac{O_{j\ell}}
{|Q_j|+|Q_\ell|-O_{j\ell}},
& j\neq\ell,\\[6pt]
0, & j=\ell.
\end{cases}
\label{eq:weighted-adjacency}
\end{equation}
Here, subscripts $j$ and $\ell$
identify the corresponding records
in the fixed observed-record ordering.
The edge set $E$ contains pairs of
distinct observed identifiers with
positive edge weights.

Jaccard weighting measures the overlap
between two records' retrieval histories
relative to their combined occurrences.
It reduces the influence of records
that appear frequently in otherwise
unrelated queries.
The zero diagonal excludes self-links,
while records never retrieved together
have zero edge weight.

\subsection{Recovering Anonymous Record Groups}
\label{sec:anonymous-groups}

\noindent\textbf{Graph normalization.}
Let $d_j=\sum_{\ell}W_{j\ell}$
denote the weighted degree of an
observed record, and define
$D=\operatorname{diag}(d_j)$.
SAGE constructs the symmetric
normalized adjacency matrix
\begin{equation}
    S=D^{-1/2}WD^{-1/2},
    \label{eq:normalized-graph}
\end{equation}
where zero-degree entries of
$D^{-1/2}$ are defined as zero.

Let $V_+\subseteq V$ denote the
observed identifiers whose graph
degrees are positive, subject to
numerical tolerance.
SAGE restricts the normalized matrix
to these records, obtaining
$S_+=S[V_+,V_+]$.
Isolated records are excluded from
spectral grouping because they provide
no graph-neighbor evidence.

\noindent\textbf{Spectral embedding.}
SAGE computes the eigenvectors
corresponding to the largest algebraic
eigenvalues of $S_+$.
Let $p$ denote the selected embedding
dimension, and let
$U\in\mathbb{R}^{|V_+|\times p}$
contain these eigenvectors.
Each row is normalized to unit length,
producing the spectral representation
$\bar U$.
These representations capture structural
relationships among records based on
their co-retrieval histories.

\noindent\textbf{Anonymous grouping.}
SAGE applies KMeans to the normalized
spectral representations to obtain
$C$ anonymous record groups:
\begin{equation}
    \boldsymbol{z}
    =\operatorname{KMeans}(\bar U,C),
    \label{eq:anonymous-grouping}
\end{equation}
where $z_j$ denotes the group
assignment of record $r_j$.
The numerical group indices are
arbitrary and do not represent
semantic categories or record labels.

Both attack variants use this spectral
framework, but their grouping
procedures serve different purposes.
SAGE-Dist uses anonymous groups
for distribution-based class assignment,
whereas SAGE-Known uses them to
aggregate evidence from known
record labels. Their specific inference
procedures are described in
Section~\ref{sec:attack}.

Records excluded from spectral
grouping receive the provisional
group marker $z_j=-1$, indicating
an unassigned group.
If the graph contains too few
non-isolated records, spectral
grouping cannot proceed.
SAGE-Dist requires successful
grouping for its initial class
assignment, whereas SAGE-Known
can still perform graph-based
label propagation.
Algorithm~\ref{alg:co-retrieval}
summarizes the graph construction
and anonymous grouping procedure.

\begin{algorithm}[t]
\DontPrintSemicolon
\small
\caption{Co-retrieval Graph and Anonymous Group Recovery}
\label{alg:co-retrieval}

\KwIn{Access-pattern leakage
$\mathcal{L}_{\mathrm{AP}}=(I_1,\ldots,I_M)$,
category count $C$, spectral dimension $p$}
\KwOut{Graph $G=(V,E,W)$ and provisional
group assignments $z$, or grouping failure}

Construct $X$ using Eq.~\eqref{eq:incidence}\;
$V\gets\bigcup_{i=1}^{M}I_i$\;
$X_V\gets$ columns of $X$ corresponding to $V$\;
$O\gets X_V^{\mathsf T}X_V$\;
Initialize $W\gets 0$\;

\ForEach{nonzero off-diagonal entry $O_{j\ell}$}{
    $W_{j\ell}\gets
    \frac{O_{j\ell}}
    {|Q_j|+|Q_\ell|-O_{j\ell}}$\;
}

$E\gets\{\{\mathrm{id}_j,\mathrm{id}_\ell\}
\mid W_{j\ell}>0,\ j\neq\ell\}$\;
$D\gets\operatorname{diag}(W\mathbf{1})$\;
$S\gets D^{-1/2}WD^{-1/2}$\;
$V_+\gets$ non-isolated observed identifiers\;
Initialize $z_j\gets-1$ for all observed records\;

\If{insufficient records for spectral grouping}{
    \Return $G=(V,E,W)$ and grouping failure\;
}

$S_+\gets S[V_+,V_+]$\;
$U\gets$ eigenvectors for the $p$ largest
algebraic eigenvalues of $S_+$\;
$\bar U\gets$ row-normalize $U$\;
$z_{V_+}\gets\operatorname{KMeans}(\bar U,C)$\;

\Return $G=(V,E,W)$ and $z$\;
\end{algorithm}

\noindent\textbf{Interpretation and limitations.}
Spectral grouping captures structural relationships
among observed records, but the resulting anonymous
groups need not coincide with semantic categories.
Mixed neighborhoods, incomplete query coverage,
and fragmented graph connections may produce
inaccurate or missing group assignments.
These groups therefore provide structural
evidence rather than recovered record labels.

\noindent\textbf{Identifier invariance.}
The co-retrieval graph depends on relationships
among returned identifiers. We formalize this property below.

\begin{proposition}[Persistent identifier-renaming invariance]
\label{prop:id-invariance}
Let $\phi$ be a fixed injective mapping on
record identifiers. Replacing every observed
access set $I_i$ with
$\phi(I_i)=\{\phi(v)\mid v\in I_i\}$
preserves the co-retrieval graph and its
edge weights up to node relabeling.
\end{proposition}

Proposition~\ref{prop:id-invariance} implies
that fixed identifier pseudonymization,
including deterministic identifier encryption,
does not eliminate the co-retrieval structure
exploited by SAGE.
For SAGE-Known, the known identifier-label
pairs must also be consistently remapped.
In contrast, identifiers that cannot be
linked across queries may disrupt the
accumulation of co-retrieval relationships.
The proof is provided in
Appendix~\ref{app:id-invariance}.

\section{Semantic Inference Attacks}
\label{sec:attack}

Section~\ref{sec:graph-construction} constructs
the co-retrieval graph $G=(V,E,W)$ and
describes how anonymous record groups are
recovered through spectral analysis.
We now present two independent attacks
that exploit this structure using different
forms of auxiliary knowledge.
SAGE-Dist uses estimated class proportions,
whereas SAGE-Known uses a small set of
known record labels.
Both attacks produce record-label predictions
and subsequently infer query categories.

\subsection{SAGE-Dist: Distribution-based Inference}
\label{sec:sage-dist}

\noindent\textbf{Auxiliary knowledge.}
SAGE-Dist uses the estimated class proportions
$\hat{\boldsymbol{\pi}}
=(\hat{\pi}_c)_{c\in\mathcal C}$,
where $\hat{\pi}_c\geq0$ and
$\sum_{c\in\mathcal C}\hat{\pi}_c=1$.
Each proportion corresponds to a named
semantic category and may be estimated
from an independent labeled dataset.
These estimates need not equal the
category proportions among records
observed by the adversary.
SAGE-Dist requires no known record labels
and makes no assumption about the
distribution of query categories.

\noindent\textbf{Distribution-guided grouping.}
SAGE-Dist obtains normalized spectral
embeddings $\bar U$ using the procedure in
Section~\ref{sec:anonymous-groups}
and partitions the non-isolated records
into $C$ anonymous groups using KMeans.
To account for clustering variability,
it considers multiple KMeans initializations
and selects a grouping based solely on
its agreement with the estimated
class proportions.

For a candidate grouping with assignments
$z_j$, the relative size of group $g$ is
\begin{equation}
    p_g=
    \frac{
    |\{\mathrm{id}_j\in V_+\mid z_j=g\}|
    }{|V_+|}.
    \label{eq:dist-group-proportion}
\end{equation}
\noindent\textbf{Naming anonymous groups.}
SAGE-Dist associates the $C$ anonymous
groups with the $C$ semantic categories
by solving
\begin{equation}
    \sigma^*=
    \arg\min_{\sigma\in\operatorname{Sym}(\mathcal C)}
    \sum_{g=0}^{C-1}|p_g-\hat{\pi}_{\sigma(g)}|,
    \label{eq:dist-matching}
\end{equation}
where $\operatorname{Sym}(\mathcal C)$
denotes all permutations of the
category indices.
The assignment is computed using
the Hungarian algorithm.

For each candidate KMeans grouping,
SAGE-Dist computes its optimal assignment
and the corresponding total discrepancy.
It retains the candidate with the
smallest discrepancy.
The selected assignment provides
initial record-label predictions
$\hat{\lambda}^{(0)}_j=\sigma^*(z_j)$
for grouped records.

\noindent\textbf{Optimal-transport refinement.}
Anonymous groups may contain records
from multiple categories, making
group-level assignments insufficient
for accurate record-label inference.
SAGE-Dist therefore refines the
initial predictions using
entropy-regularized optimal transport.

Let $\boldsymbol{\mu}_c$ denote the
spectral center of category $c$,
initialized from the named groups.
SAGE-Dist computes a nonnegative
transport plan
$\Gamma\in\mathbb R_+^{|V_+|\times C}$
by minimizing
\begin{equation}
\begin{aligned}
\min_{\Gamma\geq0}\quad
&\sum_{j,c}\Gamma_{jc}
\|\bar U_{j,:}-\boldsymbol{\mu}_c\|_2^2
+\varepsilon\sum_{j,c}
\Gamma_{jc}\log\Gamma_{jc}\\
\text{s.t.}\quad
&\sum_c\Gamma_{jc}=\frac{1}{|V_+|},
\quad \mathrm{id}_j\in V_+,\\
&\sum_j\Gamma_{jc}=\hat{\pi}_c,
\quad c\in\mathcal C.
\end{aligned}
\label{eq:dist-ot}
\end{equation}
Here, both sums over $j$ range over
records in $V_+$, and
$\varepsilon>0$ controls
entropy regularization.
The row constraints assign equal
total mass to each record, while
the column constraints align
the soft category masses with
the estimated class proportions.

SAGE-Dist solves this problem
using Sinkhorn iterations.
It assigns each record the
category with the largest
transport mass,
\begin{equation}
    \hat{\lambda}_j
    =
    \arg\max_{c\in\mathcal C}
    \Gamma_{jc},
    \label{eq:dist-ot-prediction}
\end{equation}
and updates the category centers
using the resulting predictions.
This refinement continues until
the assignments stabilize or
a stopping limit is reached.

The marginal constraints apply
to the soft transport plan.
The resulting hard predictions
are not guaranteed to match
the estimated class proportions
exactly.
Records excluded from spectral
embedding are provisionally
assigned $\hat{\lambda}_j=-1$,
where $-1$ denotes an unassigned
record label.
Their predictions are completed
in Section~\ref{sec:label-completion}.
Algorithm~\ref{alg:sage-dist}
summarizes the complete
SAGE-Dist procedure.

\begin{algorithm}[t]
\DontPrintSemicolon
\small
\caption{SAGE-Dist: Distribution-based Record Inference}
\label{alg:sage-dist}

\KwIn{Co-retrieval graph $G=(V,E,W)$,
estimated class proportions
$\hat{\boldsymbol{\pi}}$,
category count $C$,
spectral dimension $p$}
\KwOut{Provisional record labels
$\hat{\boldsymbol{\lambda}}$}

Obtain spectral embeddings $\bar U$
on $V_+$ using the adjacency matrix $W$\;

\If{spectral grouping is unavailable}{
    \Return grouping failure\;
}

\ForEach{KMeans initialization}{
    $z\gets\operatorname{KMeans}(\bar U,C)$\;
    Compute group proportions $p_g$\;
    Find the group-to-category mapping
    using Eq.~\eqref{eq:dist-matching}\;
    Retain the candidate with the
    smallest matching discrepancy\;
}

Initialize $\hat{\lambda}^{(0)}_j
\gets\sigma^*(z_j)$ for $\mathrm{id}_j\in V_+$
using the retained candidate\;

Initialize category centers
$\boldsymbol{\mu}_c$
from $\hat{\boldsymbol{\lambda}}^{(0)}$\;
$\hat{\boldsymbol{\lambda}}
\gets\hat{\boldsymbol{\lambda}}^{(0)}$\;

\ForEach{refinement iteration}{
    Compute the squared distances from
    $\bar U$ to the category centers\;
    Solve Eq.~\eqref{eq:dist-ot}
    with Sinkhorn to obtain $\Gamma$\;
    $\hat{\lambda}'_j
    \gets\arg\max_c\Gamma_{jc}$\;

    \If{$\hat{\boldsymbol{\lambda}}'
    =\hat{\boldsymbol{\lambda}}$}{
        \textbf{break}\;
    }

    $\hat{\boldsymbol{\lambda}}
    \gets\hat{\boldsymbol{\lambda}}'$\;
    Update category centers
    $\boldsymbol{\mu}_c$\;
}

Assign $\hat{\lambda}_j\gets-1$
for $\mathrm{id}_j\in V\setminus V_+$\;
\Return $\hat{\boldsymbol{\lambda}}$\;
\end{algorithm}

\noindent\textbf{Limitations and identifiability.}
Distribution-based inference depends
on the correspondence between anonymous
group sizes and category proportions.
Uneven observation coverage, imperfect
grouping, and inaccurate auxiliary
estimates may distort this relationship.
Categories with similar proportions
may also be difficult to distinguish.

Proposition~\ref{prop:dist-identifiability}
in Appendix~\ref{app:dist-identifiability}
provides a sufficient condition for
uniquely identifying the group-to-category
correspondence while accounting for
group-size discrepancies and prior
estimation errors.

\subsection{SAGE-Known: Anchor-based Inference}
\label{sec:sage-known}

\noindent\textbf{Known-record anchors.}
SAGE-Known assumes the auxiliary
identifier-label pairs $\Psi$
defined in Eq.~\eqref{eq:known-labels}.
Only records whose identifiers
belong to $V$ can serve as
anchors in the observed graph.
We denote these matched pairs by
\begin{equation}
    \Psi_V=
    \{(\mathrm{id}_j,\lambda_j)\in\Psi
    \mid \mathrm{id}_j\in V\}.
    \label{eq:observed-anchors}
\end{equation}

\noindent\textbf{Clamped label propagation.}
Define the anchor matrix
$Y\in\{0,1\}^{|V|\times C}$.
For an observed record $r_j$,
$Y_{jc}=1$ if
$(\mathrm{id}_j,c)\in\Psi_V$,
and $Y_{jc}=0$ otherwise.

Starting from $F^{(0)}=Y$,
SAGE-Known propagates anchor
information over the normalized
co-retrieval graph:
\begin{equation}
    \widetilde F^{(t+1)}
    =
    \alpha SF^{(t)}
    +(1-\alpha)Y,
    \label{eq:known-propagation}
\end{equation}
where $S$ is defined in
Eq.~\eqref{eq:normalized-graph}
and $\alpha\in(0,1)$ controls
the contribution of propagated
information.

After each iteration,
the scores of known anchors
are reset to their original
one-hot labels:
\begin{equation}
    F^{(t+1)}_{j,:}=Y_{j,:},
    \qquad
    (\mathrm{id}_j,\lambda_j)\in\Psi_V.
    \label{eq:known-clamp}
\end{equation}
All other rows retain the scores
from $\widetilde F^{(t+1)}$.
The iteration terminates upon
convergence or after a prescribed
maximum number of iterations.
Clamping ensures that known
record labels remain unchanged
during propagation.

\noindent\textbf{Class-mass normalization.}
Records in densely connected regions
may accumulate greater propagation
scores, potentially causing predictions
to favor certain categories.
To mitigate this imbalance,
SAGE-Known normalizes each
category column of the converged
score matrix $F$:
\begin{equation}
    R_{jc}=
    \begin{cases}
    \dfrac{F_{jc}}{\sum_u F_{uc}},
    &\sum_u F_{uc}>0,\\[6pt]
    0,&\text{otherwise}.
    \end{cases}
    \label{eq:known-class-mass}
\end{equation}
The sum over $u$ includes all
observed record rows.
This normalization reduces
the dominance of categories
with disproportionately large
propagation mass.

For subsequent fusion,
SAGE-Known further normalizes
each row with positive mass:
\begin{equation}
    \bar R_{jc}=
    \begin{cases}
    \dfrac{R_{jc}}{\sum_{h\in\mathcal C}R_{jh}},
    &\sum_{h\in\mathcal C}R_{jh}>0,\\
    0,&\text{otherwise}.
    \end{cases}
    \label{eq:known-row-normalization}
\end{equation}

\noindent\textbf{Group-level anchor votes.}
SAGE-Known also uses the
anonymous groups recovered
through spectral clustering.
These groups provide an
additional source of
record-label evidence
independent of propagation.

For each anonymous group $g$,
let $n_{gc}$ denote the number
of observed anchors in that
group with record label $c$,
and let $n_g=\sum_{c\in\mathcal C}n_{gc}$.
Using a smoothing parameter $s>0$,
the group-level vote for record
$r_j$ and category $c$ is
\begin{equation}
H_{jc}=
\begin{cases}
\displaystyle
\frac{n_{z_jc}+s}
     {n_{z_j}+sC},
&z_j\geq0\text{ and }n_{z_j}>0,\\[6pt]
0,&\text{otherwise}.
\end{cases}
\label{eq:known-group-vote}
\end{equation}

The vote reflects the distribution
of known record labels within
an anonymous group.
Smoothing prevents groups
containing few anchors from
producing excessively confident
scores.
Groups without known anchors
contribute no votes.
The same applies to records
without a group assignment.

\noindent\textbf{Score fusion and prediction.}
SAGE-Known combines the normalized
propagation scores with the
group-level anchor votes:
\begin{equation}
    T_{jc}
    =(1-\gamma)\bar R_{jc}
    +\gamma H_{jc},
    \label{eq:known-fusion}
\end{equation}
where $\gamma\in[0,1]$
controls the contribution
of group-level evidence.

The predicted record label is
\begin{equation}
    \hat{\lambda}_j=
    \arg\max_{c\in\mathcal C}T_{jc},
    \label{eq:known-prediction}
\end{equation}
provided the maximum score is
positive.
Records receiving no evidence
are provisionally assigned
$\hat{\lambda}_j=-1$.
Known anchors retain their
supplied labels regardless
of the fused scores.

If spectral grouping is
unavailable, SAGE-Known
uses the class-mass-normalized
propagation scores $R$
alone for prediction.
Algorithm~\ref{alg:sage-known}
summarizes SAGE-Known.

\begin{algorithm}[t]
\DontPrintSemicolon
\small
\caption{SAGE-Known: Anchor-based Record Inference}
\label{alg:sage-known}

\KwIn{Co-retrieval graph $G=(V,E,W)$,
observed anchors $\Psi_V$,
category count $C$,
optional group assignments $z$}
\KwOut{Provisional record labels
$\hat{\boldsymbol{\lambda}}$}

Construct one-hot anchor matrix $Y$
from $\Psi_V$\;
$F\gets Y$\;
Use the normalized adjacency matrix $S$
from Eq.~\eqref{eq:normalized-graph}\;

\Repeat{convergence or iteration limit}{
    $F'\gets\alpha SF+(1-\alpha)Y$\;
    Reset anchor rows of $F'$ to $Y$\;
    $F\gets F'$\;
}

Normalize category columns of $F$
to obtain $R$\;

\eIf{anonymous groups $z$ are available}{
    Row-normalize $R$ to obtain $\bar R$\;
    Compute smoothed group votes $H$
    from $z$ and $\Psi_V$\;
    $T\gets(1-\gamma)\bar R+\gamma H$\;
}{
    $T\gets R$\;
}

Assign $\hat{\lambda}_j\gets
\arg\max_cT_{jc}$
for positive-score rows\;
Assign $\hat{\lambda}_j\gets-1$
for zero-score rows\;
Restore all known anchor labels\;

\Return $\hat{\boldsymbol{\lambda}}$\;
\end{algorithm}

\subsection{Label Completion and Query Inference}
\label{sec:label-completion}

\noindent\textbf{Completing record predictions.}
Both attack variants may leave
some observed records with
the provisional label $-1$,
which indicates an unassigned
prediction.

SAGE first attempts to label
these records through iterative
weighted neighbor voting.
In each round, an unassigned
record with at least one
labeled neighbor receives
the category supported by
the largest total edge weight:
\begin{equation}
    \hat{\lambda}_j
    \gets
    \arg\max_{c\in\mathcal C}
    \sum_{\substack{
    \ell:\,\mathrm{id}_\ell\in V\\
    \hat{\lambda}_\ell=c
    }}W_{j\ell}.
    \label{eq:neighbor-completion}
\end{equation}

Only predictions available
at the beginning of a round
contribute to that round's
votes, making the updates
independent of processing order.
The procedure repeats until
no further records can be
assigned through neighbors.

Remaining unassigned records
receive the global majority
category among the predictions
available before completion.
If no such predictions exist,
SAGE uses category index $0$
as a fallback.

\noindent\textbf{Inferring query categories.}
After completing record-label
predictions, SAGE infers
the category of each observed
query by majority voting
over its retrieved records:
\begin{equation}
    \hat{\ell}_i
    =
    \arg\max_{c\in\mathcal C}
    \sum_{\mathrm{id}_j\in I_i}
    \mathbf{1}[\hat{\lambda}_j=c].
    \label{eq:query-category-vote}
\end{equation}

This procedure applies to
both SAGE-Dist and SAGE-Known.
If no retrieved record has
a valid prediction, the query
remains unresolved with
$\hat{\ell}_i=-1$.
Under the default completion
procedure, every observed
record receives a prediction.

Query-category inference does
not require all retrieved
records to share the same
record label.
The inferred category may
differ from the query's
true category, particularly
when retrieval crosses
category boundaries.
SAGE estimates the query
category without reconstructing
its plaintext content.

\noindent\textbf{Scope of inference.}
SAGE produces record-label
predictions only for identifiers
in the observed set $V$,
while query-category predictions
cover the $M$ observed queries.
The accuracy of these predictions
depends on the co-retrieval
structure, the available
auxiliary knowledge, and the
observed query workload.
We evaluate these factors in
Section~\ref{sec:evaluation}.

\section{Evaluation}
\label{sec:evaluation}

We evaluate semantic leakage from access patterns
across four datasets. We first examine co-retrieval
structure and attack effectiveness, then assess
the impact of algorithmic components, auxiliary
knowledge, and observation volume. Finally, we
study robustness under real encrypted retrieval.

\subsection{Experimental Setup}
\label{sec:experimental-setup}

\noindent\textbf{Datasets and embeddings.}
We evaluate four datasets spanning text and image retrieval:
AG News, Yahoo Answers, MedMCQA, and CIFAR-100, with 4, 10,
21, and 100 semantic categories, respectively. We encode
text using BGE-base-en-v1.5 (768 dimensions) and images
using OpenCLIP ViT-B/32 (512 dimensions). All embeddings
are $\ell_2$-normalized. Database, query, and auxiliary
pools are disjoint, and sampling is performed without
replacement. Ground-truth labels are used only for
evaluation and the specified auxiliary knowledge.

\noindent\textbf{Retrieval workloads.}
Unless otherwise stated, we observe $M=2{,}000$ queries,
each returning the exact inner-product top-$k$ records
with $k=100$. Database sizes depend on the dataset.
We use Zipf class distributions with exponents $1.0$
for AG News and Yahoo Answers and $0.5$ for CIFAR-100,
while MedMCQA retains its natural class distribution.
Queries are sampled uniformly from the disjoint query
pool without imposing a class distribution matched
to the database. All attacks use the returned record
identifiers rather than plaintext embeddings.

\noindent\textbf{Auxiliary knowledge and metrics.}
Both attacks know the number of semantic categories.
SAGE-Dist receives estimated class proportions from
a simulated independent auxiliary sample of 1,000
labels drawn according to the database's class
distribution, without any known target-record labels.
SAGE-Known receives labels of 1\% of randomly selected
database records, chosen independently of the queries.
Only known records appearing in the observed access
patterns can serve as graph anchors. We report
record-label accuracy and macro-F1 over observed
records excluding disclosed labels, and query-category
accuracy over the observed queries. Main results are
averaged over ten independent runs, with standard
deviations reported for record-label accuracy.

\noindent\textbf{Implementation.}
We implement SAGE in Python using NumPy, SciPy, and
scikit-learn. Both attacks construct Jaccard-weighted
co-retrieval graphs. Dataset-specific attack parameters
and further implementation details are provided in
Appendix~\ref{app:dataset-statistics}. Experiments
are conducted on a 16-core 2.20\,GHz Intel Xeon
machine with 64\,GB RAM.

\subsection{Evidence of Semantic Neighborhood Leakage}
\label{sec:snl-structure}

We first examine whether the observed access
patterns retain semantic structure before
applying either inference attack.
For this analysis, ground-truth record labels
are used only to evaluate co-retrieval
relationships and are not available
to the adversary.

\begin{figure}[t]
    \centering
    \includegraphics[width=0.8\columnwidth]
    {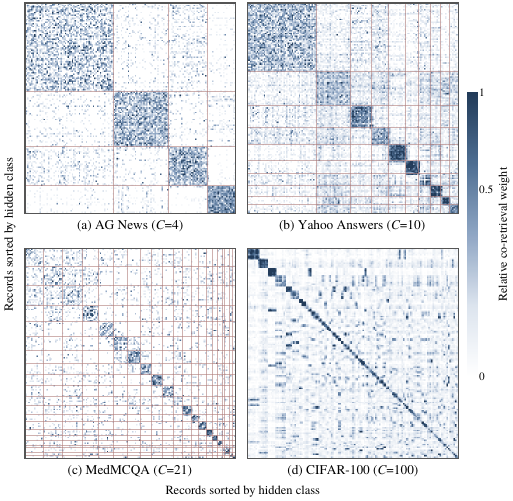}
    \caption{Record-level co-retrieval structure
    across four datasets.
    Darker cells indicate stronger pairwise
    Jaccard co-retrieval weights.
    Category boundaries are marked by
    thin red lines where visible.}
    \label{fig:class-pair-heatmap}
\end{figure}

\noindent\textbf{Semantic structure.}
Figure~\ref{fig:class-pair-heatmap}
visualizes the record-level co-retrieval
graph, with records ordered by their
ground-truth semantic categories.
The diagonal block structures indicate
stronger co-retrieval relationships
among records from the same category,
while off-diagonal connections reveal
overlaps across categories.

Table~\ref{tab:snl-structure}
quantifies this effect.
Across the four datasets, mean
within-category Jaccard weights are
5.95--25.29 times the corresponding
between-category weights.
Both averages include zero-weight
record pairs.
These results demonstrate a
statistical association between
observable co-retrieval relationships
and hidden semantic categories.

\begin{table}[t]
\centering
\caption{Semantic structure in the observed
co-retrieval graph over ten runs.
}
\label{tab:snl-structure}
\small
\setlength{\tabcolsep}{4pt}
\resizebox{0.85\columnwidth}{!}{%
\begin{tabular}{lrrr}
\toprule
Dataset & $C$ &
\shortstack{Observation\\coverage (\%)} &
\shortstack{Within/between\\weight ratio} \\
\midrule
AG News       & 4   & 84.1 & $8.58\pm0.14$ \\
Yahoo Answers & 10  & 88.4 & $5.95\pm0.09$ \\
MedMCQA       & 21  & 77.7 & $10.13\pm0.08$ \\
CIFAR-100     & 100 & 98.5 & $25.29\pm0.48$ \\
\bottomrule
\end{tabular}
}
\end{table}

\noindent\textbf{Statistical significance.}
We perform a permutation test by randomly
shuffling record labels while keeping
the co-retrieval graph and category
counts fixed.
For each dataset, none of the 500
permutations in any of the ten runs
produces a within/between ratio
at least as large as the observed value.
The resulting permutation-test
$p$-value is $1/501\approx0.002$
under the plus-one correction.

These findings establish a statistical
association between observable
co-retrieval relationships and
semantic categories.
However, structural separation alone
does not guarantee that an adversary
can assign the correct category names.
The next section evaluates how SAGE
converts this structural evidence into
record-label and query-category predictions.

\subsection{Effectiveness of SAGE}
\label{sec:attack-effectiveness}

\begin{table*}[t]
\centering
\caption{Record-label and query-category inference across four
datasets. Record, F1, and Query are percentages averaged over
ten runs. Grouping ARI, NMI, and purity are unitless.
Oracle alignment reports the best one-to-one mapping
using ground-truth labels, shown as spectral / Dist grouping
(\%). Bold marks the highest attack mean within each
auxiliary-knowledge block.}
\label{tab:sage-main}
\scriptsize
\setlength{\tabcolsep}{2.0pt}
\renewcommand{\arraystretch}{1.10}
\begin{tabular*}{0.95\textwidth}{
@{\extracolsep{\fill}}l*{12}{c}@{}
}
\toprule
 & \multicolumn{3}{c}{AG News}
 & \multicolumn{3}{c}{Yahoo Answers}
 & \multicolumn{3}{c}{MedMCQA}
 & \multicolumn{3}{c}{CIFAR-100} \\
\cmidrule(lr){2-4}
\cmidrule(lr){5-7}
\cmidrule(lr){8-10}
\cmidrule(lr){11-13}
Method
 & Record & F1 & Query
 & Record & F1 & Query
 & Record & F1 & Query
 & Record & F1 & Query \\
\midrule
\multicolumn{13}{l}{
\textit{Category count or estimated class proportions}} \\
Random
 & 25.0 & 23.6 & 26.2
 & 10.0 & 8.7 & 9.9
 & 4.7 & 4.4 & 4.5
 & 1.0 & 0.9 & 1.0 \\
Prior-Sampling
 & 31.6 & 25.1 & 25.2
 & 17.5 & 10.1 & 9.7
 & 6.2 & 4.7 & 8.5
 & 1.5 & \textbf{1.0} & \textbf{1.1} \\
Prior-Majority
 & 45.1 & 15.6 & 25.2
 & 31.8 & 4.8 & 9.7
 & \textbf{9.6} & 0.8 & \textbf{9.5}
 & \textbf{5.3} & 0.1 & 1.0 \\
SAGE-Dist
 & \textbf{66.9} & \textbf{66.8} & \textbf{68.1}
 & \textbf{36.1} & \textbf{19.9} & \textbf{20.5}
 & 7.3 & \textbf{6.0} & 7.5
 & 1.9 & 0.9 & 0.8 \\
\midrule
\multicolumn{13}{l}{\textit{1\% known record labels}} \\
Nearest-known
 & 79.8 & 78.8 & 83.7
 & 57.1 & 49.5 & 58.3
 & 40.1 & 37.9 & 47.4
 & 27.5 & \textbf{14.6} & 19.3 \\
Louvain-known
 & 79.1 & 78.1 & 80.9
 & 59.7 & 50.8 & 57.0
 & 38.3 & 34.9 & 43.0
 & 22.9 & 7.0 & 12.3 \\
Standard-LP
 & 51.5 & 26.4 & 31.7
 & 31.8 & 4.9 & 9.7
 & 36.5 & 27.8 & 38.7
 & 12.9 & 4.1 & 5.1 \\
Label-Spreading
 & 70.5 & 57.8 & 59.9
 & 32.8 & 6.4 & 9.9
 & 43.9 & 36.6 & 48.2
 & 18.4 & 8.1 & 7.7 \\
SAGE-Known
 & \textbf{84.3} & \textbf{84.1} & \textbf{87.4}
 & \textbf{64.9} & \textbf{57.9} & \textbf{63.6}
 & \textbf{47.0} & \textbf{44.5} & \textbf{52.5}
 & \textbf{27.8} & 14.5 & \textbf{20.3} \\
\midrule
\multicolumn{13}{l}{
\textit{Grouping diagnostics (ground truth used only for evaluation)}} \\
Metric
 & ARI & NMI & Purity
 & ARI & NMI & Purity
 & ARI & NMI & Purity
 & ARI & NMI & Purity \\
Spectral grouping
 & .408 & .497 & .749
 & .223 & .383 & .606
 & .179 & .314 & .421
 & .267 & .607 & .488 \\
Dist-selected grouping
 & .461 & .494 & .767
 & .376 & .384 & .600
 & .132 & .274 & .365
 & .138 & .551 & .423 \\
Oracle alignment
 & \multicolumn{3}{c}{64.9 / 75.4}
 & \multicolumn{3}{c}{45.0 / 57.6}
 & \multicolumn{3}{c}{36.7 / 31.6}
 & \multicolumn{3}{c}{41.3 / 33.8} \\
\bottomrule
\end{tabular*}
\end{table*}

We evaluate record-label and query-category inference under
SAGE-Dist's estimated class-proportion setting and SAGE-Known's
1\% known-label setting. Table~\ref{tab:sage-main} compares each
attack with methods using the same auxiliary knowledge.
The two auxiliary-knowledge settings are evaluated separately.

\noindent\textbf{SAGE-Dist Effectiveness.}
With estimated class proportions, SAGE-Dist achieves 66.9\% record-label
accuracy on AG News and 36.1\% on Yahoo Answers, exceeding
Prior-Majority by 21.8 and 4.3 percentage points, respectively.
On MedMCQA and CIFAR-100, SAGE-Dist reaches only 7.3\% and
1.9\%, below Prior-Majority at 9.6\% and 5.3\%.
This stem from the semantic similarity among MedMCQA
categories and the large number of categories in CIFAR-100. Thus, the class-associated co-retrieval
structure identified in Section~\ref{sec:snl-structure}
does not guarantee accurate semantic label recovery.

\noindent\textbf{SAGE-Known Effectiveness.}
With 1\% known record labels, SAGE-Known achieves 84.3\%,
64.9\%, 47.0\%, and 27.8\% record-label accuracy.
Beyond the access-pattern heuristics Nearest-known and
Louvain-known, we include two classical graph-based
semi-supervised baselines, Standard-LP and Label-Spreading,
without class-mass normalization or spectral group voting.
SAGE-Known consistently outperforms the strongest
independent baseline in mean record-label accuracy
across all four datasets.
We isolate the contribution of group-level voting from the
propagation component in Section~\ref{sec:auxiliary-knowledge}.

\noindent\textbf{Query-category inference.}
SAGE-Known achieves query-category accuracies of 87.4\%,
63.6\%, 52.5\%, and 20.3\%, respectively, exceeding the
independent known-label baselines in Table~\ref{tab:sage-main}.
SAGE-Dist reaches 68.1\% and 20.5\% on AG News and Yahoo
Answers, but remains weak on MedMCQA and CIFAR-100.
Because query categories are predicted by voting over retrieved
record labels, their accuracy depends on the composition of
retrieval results as well as record-label recovery.
Implementation details of the comparison methods appear in
Appendix~\ref{app:attack-comparisons}.

\noindent\textbf{Grouping versus semantic identification.}
Table~\ref{tab:sage-main} reports grouping diagnostics
to distinguish structural recovery from semantic
label inference. On CIFAR-100, spectral grouping
achieves an NMI of 0.607, while the grouping selected
by SAGE-Dist achieves an NMI of 0.551 and an oracle
alignment accuracy of 33.8\%. Nevertheless,
SAGE-Dist reaches only 1.9\% record-label accuracy.
A similar gap appears on MedMCQA, where the
Dist-selected grouping achieves 31.6\% oracle
alignment accuracy but only 7.3\% actual accuracy.

These results indicate that anonymous groups can
retain semantic information even when the attack
fails to identify record categories accurately.
However, oracle alignment uses ground-truth labels
unavailable to the adversary, and the gap also
reflects subsequent optimal-transport refinement.
Thus, structural recovery alone does not guarantee
successful semantic identification.

\subsection{Ablation Studies}
\label{sec:auxiliary-knowledge}

\begin{figure}[t]
    \centering
    \includegraphics[width=\columnwidth]{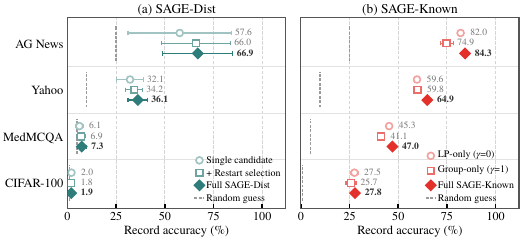}
    \caption{Component ablations of (a) SAGE-Dist and
    (b) SAGE-Known. Dist adds restart selection and Sinkhorn
    refinement. Known compares LP-only ($\gamma=0$),
    Group-only ($\gamma=1$), and full fusion.}
    \label{fig:sage-ablation}
\end{figure}

We examine whether the main components of SAGE-Dist and
SAGE-Known contribute to record-label inference under their
respective auxiliary-knowledge settings. Within each attack,
the compared variants share the same access patterns and
auxiliary knowledge. Figure~\ref{fig:sage-ablation} reports
mean record-label accuracy and standard deviation over ten
runs. Exact results and fusion-weight sensitivity are given
in Appendix~\ref{app:sage-fusion}.

\noindent\textbf{SAGE-Dist components.}
Figure~\ref{fig:sage-ablation}(a) compares a single KMeans
candidate named by class-proportion matching with two
successive additions: selection among six candidates and
Sinkhorn-based record-level refinement. Both additions use
the same spectral embedding dimension and auxiliary prior.
On AG News, mean accuracy increases from 57.6\% to 66.0\%
and then 66.9\%. Yahoo Answers improves from 32.1\% to
34.2\% and then 36.1\%, while MedMCQA shows smaller gains.
On CIFAR-100, the full method reaches 1.9\%. These findings
show that the added stages can improve category assignment,
but they cannot reliably resolve category-name ambiguity
from class proportions in every dataset.

\noindent\textbf{SAGE-Known components.}
Figure~\ref{fig:sage-ablation}(b) compares clamped label
propagation with class-mass normalization (LP-only,
$\gamma=0$), spectral group-level anchor voting without
propagation scores (Group-only, $\gamma=1$), and their
fusion in SAGE-Known. The full method achieves the highest
mean record-label accuracy among these variants on all
four datasets. Relative to LP-only, its gains are 2.3,
5.2, 1.7, and 0.3 percentage points on AG News, Yahoo
Answers, MedMCQA, and CIFAR-100, respectively.
Group-only also underperforms the fused method, indicating
that group-level votes are complementary to propagation
rather than a sufficient replacement. The full $\gamma$
sweep in Appendix~\ref{app:sage-fusion} further shows that
the contribution of group voting depends on its weight.

\begin{figure*}[t]
    \centering
    \includegraphics[width=\textwidth]{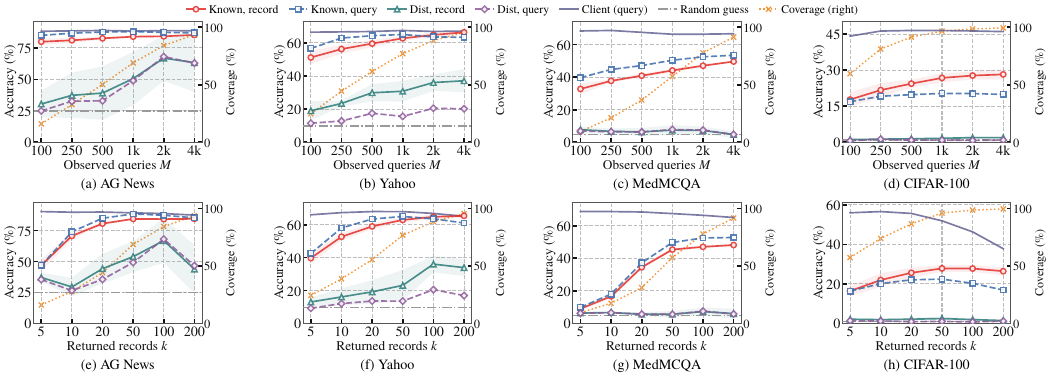}
    \caption{Impact of observation workload on SAGE
    across four datasets. Panels (a)--(d) vary the
    number of observed queries $M$ with $k=100$,
    while panels (e)--(h) vary the retrieval size $k$
    with $M=2000$. Curves show record-label and
    query-category accuracy for both attacks,
    client query-category accuracy, and random
    guessing. The right axis shows observed-record
    coverage. Results are averaged over ten runs.}
    \label{fig:sage-workload}
\end{figure*}

\subsection{Observation and Auxiliary Knowledge}
\label{sec:observation-auxiliary}

We examine how attack effectiveness depends on the
number of observed queries $M$, retrieval size $k$,
and known-label budget. All results are averaged
over ten runs. Since varying $M$ or $k$ changes the
observed record set, record-label accuracy is
evaluated over observed records excluding disclosed
labels.

\noindent\textbf{Observation volume.}
Figure~\ref{fig:sage-workload}(a)--(d) shows that
persistent observation benefits SAGE-Known.
As $M$ increases from 100 to 4,000, its
record-label accuracy rises from 80.0\% to
85.3\% on AG News, 51.2\% to 66.4\% on Yahoo
Answers, 32.8\% to 49.6\% on MedMCQA, and
17.8\% to 28.3\% on CIFAR-100.
SAGE-Dist also benefits from additional
observations on AG News and Yahoo Answers,
but shows no consistent improvement on
MedMCQA and CIFAR-100. More observations
provide richer co-retrieval evidence, but
do not necessarily resolve ambiguous
group-to-category assignments.

\noindent\textbf{Retrieval size.}
Figure~\ref{fig:sage-workload}(e)--(h) examines
$k\in\{5,10,20,50,100,200\}$.
Increasing $k$ initially improves SAGE-Known,
but its accuracy eventually saturates or
declines on some datasets.
SAGE-Dist also exhibits non-monotonic behavior,
particularly on AG News and Yahoo Answers.
Although larger retrieval sets expose more
records and co-retrieval relationships,
they may also introduce additional
cross-category connections that weaken
semantic separability.

\begin{figure*}[t]
    \centering
    \includegraphics[width=\textwidth]{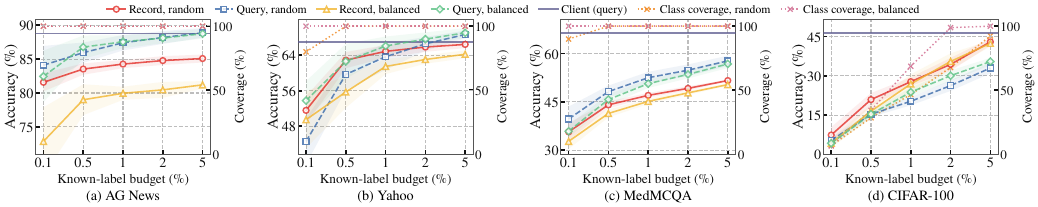}
    \caption{Impact of known-label budget on SAGE-Known.
    We vary the known-label fraction from 0.1\% to
    5\% under random and class-balanced anchor
    selection. Curves report record-label and
    query-category accuracy, client query
    accuracy, and known-class coverage.
    Class-balanced selection is an idealized
    diagnostic using ground-truth labels.
    Results are averaged over ten runs.}
    \label{fig:sage-known-budget}
\end{figure*}

\noindent\textbf{Known-label budget.}
Figure~\ref{fig:sage-known-budget} shows that increasing
the fraction of randomly known labels from 0.1\%
to 5\% improves SAGE-Known's record-label accuracy
from 81.6\% to 85.1\% on AG News, 51.6\% to
66.4\% on Yahoo Answers, 35.7\% to 51.6\%
on MedMCQA, and 7.4\% to 43.0\% on CIFAR-100.
The largest gain occurs on CIFAR-100, where
usable known-class coverage increases from
6.8\% to 92.4\%. Class-balanced selection
improves category coverage in some settings
but does not consistently improve inference
accuracy. Thus, both anchor availability and
category coverage affect semantic recovery.

\subsection{Robustness in Encrypted Vector Retrieval}
\label{sec:retrieval-variants}

We evaluate SAGE with Quiver's encrypted vector retrieval using
four retrieval settings. \emph{Exact} performs local plaintext
exact top-$k$ search. \emph{Quiver Plain} uses the server's
plaintext HNSW index. \emph{Simulated DCPE} applies local
scale-and-perturb noise followed by exact retrieval.
\emph{Quiver DCPE} combines client-side DCPE encryption
with Quiver's HNSW retrieval. In all settings, the adversary
uses only the returned record identifiers. We use
$M=2{,}000$ queries and $k=100$ throughout.
The full protocol is described in
Appendix~\ref{app:quiver-protocol}.

\begin{table}[t]
\centering
\caption{Record-label accuracy and retrieval Recall@100
under exact search and Quiver DCPE at $\beta=1.6$ (\%).
All values are means over ten runs. Recall is measured
against exact plaintext top-100 retrieval.}
\label{tab:sage-quiver-main}
\scriptsize
\setlength{\tabcolsep}{2.6pt}
\renewcommand{\arraystretch}{1.10}
\resizebox{0.85\columnwidth}{!}{%
\begin{tabular}{lccccc}
\toprule
 & \shortstack{Recall@100}
 & \multicolumn{2}{c}{SAGE-Dist}
 & \multicolumn{2}{c}{SAGE-Known} \\
\cmidrule(lr){3-4}\cmidrule(lr){5-6}
Dataset & DCPE & Exact & DCPE & Exact & DCPE \\
\midrule
AG News & 76.9 & 66.9 & 61.7 & 84.3 & 84.2 \\
Yahoo & 74.8 & 36.1 & 32.8 & 64.9 & 64.7 \\
MedMCQA & 76.9 & 7.3 & 5.4 & 47.0 & 46.9 \\
CIFAR-100 & 68.8 & 1.9 & 1.9 & 27.8 & 27.1 \\
\bottomrule
\end{tabular}%
}
\end{table}

\noindent\textbf{Attack effectiveness.}
At $\beta=1.6$, Quiver DCPE achieves only
68.8--76.9\% Recall@100, yet SAGE-Known maintains
record-label accuracies of 84.2\%, 64.7\%, 46.9\%,
and 27.1\% across the four datasets, close to its
exact-search results. Thus, retrieval errors do not
necessarily eliminate the co-retrieval relationships
exploited by the attack. SAGE-Dist is less stable,
with 61.7\%, 32.8\%, 5.4\%, and 1.9\% accuracy,
respectively. Its limited performance on MedMCQA
and CIFAR-100 also occurs under exact retrieval.

\noindent\textbf{Perturbation sensitivity.}
We further vary the DCPE perturbation strength
$\beta\in\{0.02,0.1,0.4,1.6\}$. As $\beta$ increases,
Quiver DCPE Recall@100 decreases from approximately
99\% to 68.8--76.9\%, while SAGE-Known accuracy
changes only slightly. SAGE-Dist shows greater
variation without a consistent trend.
Results for both Quiver DCPE and the simulated
perturbation are given in
Appendix~\ref{app:quiver-protocol}.

\noindent\textbf{Record linkability.}
We further examine idealized identifier rotation,
where record pseudonyms cannot be linked across
query windows. With windows of 50--500 queries,
SAGE-Dist falls below the majority baseline on
AG News and Yahoo Answers. SAGE-Known also degrades
substantially when known-label anchors can only
be recognized in the first window, whereas access
to anchors in every window preserves much of its
accuracy. Unlike Quiver DCPE, which retains linkable
record identifiers, this experiment removes
cross-window linkability by assumption.
It quantifies the importance of persistent
record linkage rather than demonstrating a
deployable defense. Further results are given
in Appendix~\ref{app:rotation}.

\section{Discussion and Limitations}

\noindent\textbf{Scope of the attack.}
SAGE targets encrypted vector-search systems
with record identifiers that remain linkable
across queries. It requires neither plaintext
embeddings or queries nor similarity scores
or retrieval rankings. Its effectiveness
depends on the co-retrieval relationships
exposed by legitimate workloads, a known
semantic taxonomy, and limited auxiliary
knowledge. SAGE infers record labels and
query categories rather than reconstructing
plaintext contents. Systems that hide
retrieval accesses fall outside our threat model.

\noindent\textbf{Limitations of semantic inference.}
Recovering anonymous record groups does not
guarantee identifying their semantic categories.
SAGE-Dist performs better on AG News and Yahoo
Answers but remains limited on MedMCQA and
CIFAR-100, where ambiguous class proportions,
mixed neighborhoods, and imperfect grouping
complicate category assignment. SAGE-Known
achieves stronger results but depends on the
availability and category coverage of known
labels. Both variants may be affected by
changes in query distributions, embeddings,
and retrieval workloads. Our findings do not
imply universal label recoverability.

\noindent\textbf{Deployment implications and defenses.}
Experiments with Quiver DCPE demonstrate that
semantic inference remains possible under
encrypted approximate retrieval. Fixed identifier
pseudonymization does not prevent co-retrieval
analysis when identifiers remain linkable.
Identifier rotation may reduce accumulated
leakage if cross-window linkage is prevented,
although other metadata may still enable
record linkage. Such defenses can introduce
additional system costs. These findings
motivate protecting semantic relationships
alongside encrypted vector contents.

\section{Conclusion}

We identify semantic neighborhood leakage
in encrypted vector search, showing that
co-retrieval relationships can expose
sensitive semantic information even when
record contents and embeddings remain protected.
We develop SAGE-Dist and SAGE-Known to infer
record labels and query categories using
distinct forms of auxiliary knowledge.
Experiments across four datasets and encrypted
retrieval settings demonstrate the feasibility
and limitations of these attacks. Our findings
highlight the need to protect semantic
relationships beyond encrypting vector contents.


% % use section* for acknowledgment
% \ifCLASSOPTIONcompsoc
%   % The Computer Society usually uses the plural form
%   \section*{Acknowledgments}
% \else
%   % regular IEEE prefers the singular form
% \fi

% The authors would like to thank...no one



%\bibliographystyle{IEEEtran}


% \end{thebibliography}

\bibliographystyle{IEEEtran}
\bibliography{refs}

\appendices

\section{Record-Identifier Exposure in Practice}
\label{app:identifier_exposure}

Practical vector search systems associate retrieved
records with identifiers. Figure~\ref{fig:industrial-ids}
illustrate identifier-bearing retrieval interfaces
in Milvus, Qdrant, Weaviate, and Quiver.

\begin{figure*}[t]
\centering

\begin{minipage}[t]{0.48\textwidth}
\vspace{0pt}
\begin{casebox}{Case 1: Milvus}
hits = client.search(
    collection_name="quick_setup",
    data=[q], limit=10
)
ids = [h["id"] for h in hits[0]]
\end{casebox}
\end{minipage}
\hfill
\begin{minipage}[t]{0.48\textwidth}
\vspace{0pt}
\begin{casebox}{Case 2: Qdrant}
hits = client.query_points(
    collection_name="test_collection",
    query=q, limit=10
).points
ids = [p.id for p in hits]
\end{casebox}
\end{minipage}

\vspace{2mm}

\begin{minipage}[t]{0.48\textwidth}
\vspace{0pt}
\begin{casebox}{Case 3: Weaviate}
results = collection.query.near_vector(
    near_vector=q, limit=10
)
ids = [o.uuid for o in results.objects]
\end{casebox}
\end{minipage}
\hfill
\begin{minipage}[t]{0.48\textwidth}
\vspace{0pt}
\begin{casebox}{Case 4: Quiver DCPE}
enc_q = cipher.encrypt_query(q)
hits = db.search(
    "vault", enc_q, k=10
)
print(hits)
\end{casebox}
\end{minipage}

\caption{Examples of identifier-bearing retrieval
interfaces in practical vector search systems.
The snippets assume initialized clients and
query inputs.}
\label{fig:industrial-ids}
\end{figure*}

\noindent\textbf{Server-side observation in Quiver.}
Unlike the preceding client-interface examples,
our Quiver DCPE experiment examines identifiers
in serialized retrieval responses at the
server-egress observation point.
The server performs approximate similarity search
over encrypted vectors and selects the returned
records without accessing plaintext embeddings.

The following excerpt contains returned identifiers
from two queries in our AG News experiment:

\begin{verbatim}
query_index: 0
returned_ids:
  7011d23e7b324832b882fb13e4dae0bf
  7a98edf6049244eaa58ce8aba251ded0

query_index: 1
returned_ids:
  e3f78c99d8654bd5a01de8b75c5a9fce
  69ebbe2e9f34474dabe3844b821e5b0e
\end{verbatim}

Only the first two identifiers per query are shown.
Although these identifiers are opaque, their
cross-query linkability allows the adversary
to construct co-retrieval relationships.
This observation illustrates the access-pattern
leakage assumed by SAGE without requiring
plaintext embeddings, queries, or record labels.

\section{Theoretical Properties of SAGE}
\label{app:theory}

This appendix establishes two properties
relevant to SAGE. The first shows that
persistent identifier renaming preserves
the co-retrieval structure. The second
provides a sufficient condition for
identifying semantic classes from anonymous
group sizes and estimated class proportions.

\subsection{Invariance to Persistent ID Renaming}
\label{app:id-invariance}

\noindent\textbf{Proof of
Proposition~\ref{prop:id-invariance}.}
Let $\phi$ be a fixed injective mapping on record
identifiers, and let $I_i'=\phi(I_i)$.
For every observed record $\mathrm{id}_j$,
\begin{equation}
\begin{aligned}
Q'(\phi(\mathrm{id}_j))
&=\{i\mid\phi(\mathrm{id}_j)\in I_i'\}\\
&=\{i\mid\mathrm{id}_j\in I_i\}
=Q_j.
\end{aligned}
\end{equation}
The second equality follows from the injectivity
of $\phi$. Thus, for any two observed records,
their Jaccard weight remains unchanged:
\begin{equation}
\frac{|Q'(\phi(\mathrm{id}_j))
\cap Q'(\phi(\mathrm{id}_\ell))|}
{|Q'(\phi(\mathrm{id}_j))
\cup Q'(\phi(\mathrm{id}_\ell))|}
=
\frac{|Q_j\cap Q_\ell|}
{|Q_j\cup Q_\ell|}
=W_{j\ell}.
\end{equation}
Consequently, the renamed graph is isomorphic
to $G$ and preserves its weighted structure.
For SAGE-Known, the auxiliary identifier-label
pairs must also be consistently renamed.
The result requires persistent identifier
linkability and does not apply to independently
unlinkable identifiers across queries.

\subsection{Identifiability from Class Proportions}
\label{app:dist-identifiability}

We establish a sufficient condition for
SAGE-Dist to correctly associate anonymous
groups with semantic categories.
Suppose the $C$ groups admit a correct
one-to-one correspondence $\sigma_0$.
Let $\boldsymbol{\pi}$ denote the true
class proportions, which are unknown to
the attacker. Define
\begin{equation}
\begin{aligned}
\epsilon_{\mathrm{grp}}
&=\max_g|p_g-\pi_{\sigma_0(g)}|,\\
\epsilon_{\mathrm{prior}}
&=\|\hat{\boldsymbol{\pi}}
-\boldsymbol{\pi}\|_\infty,\\
\Delta_\pi
&=\min_{c\neq c'}|\pi_c-\pi_{c'}|.
\end{aligned}
\label{eq:dist-error-definitions}
\end{equation}

\begin{proposition}[Identifiability of class assignment]
\label{prop:dist-identifiability}
If a correct bijection $\sigma_0$ exists and
\begin{equation}
\Delta_\pi>
2(\epsilon_{\mathrm{grp}}
+\epsilon_{\mathrm{prior}}),
\label{eq:dist-identifiability-condition}
\end{equation}
then $\sigma_0$ uniquely minimizes the
assignment objective in
Eq.~\eqref{eq:dist-matching}.
\end{proposition}

\begin{proof}
For each group $g$, let $c_0=\sigma_0(g)$.
By the triangle inequality,
\begin{equation}
|p_g-\hat{\pi}_{c_0}|
\leq\epsilon_{\mathrm{grp}}
+\epsilon_{\mathrm{prior}}.
\end{equation}
For any incorrect class $c\neq c_0$,
the reverse triangle inequality gives
\begin{equation}
|p_g-\hat{\pi}_c|
\geq\Delta_\pi-\epsilon_{\mathrm{grp}}
-\epsilon_{\mathrm{prior}}.
\end{equation}
Under the stated condition,
\begin{equation}
|p_g-\hat{\pi}_{c_0}|
<
|p_g-\hat{\pi}_c|.
\end{equation}
Thus, every group strictly prefers its
correct class. Since $\sigma_0$ is a
bijection, any other feasible assignment
has strictly greater total cost.
Therefore, $\sigma_0$ is the unique
minimizer.
\end{proof}

\noindent\textbf{Scope.}
The condition guarantees only the
uniqueness of the group-to-class assignment
when a correct correspondence exists.
It does not guarantee accurate spectral
grouping, record-label recovery within
mixed groups, or improvement from
optimal-transport refinement.

\section{Additional Experimental Details and Results}
\label{app:additional-evaluation}

\subsection{Dataset and Attack Configuration}
\label{app:dataset-statistics}

\noindent\textbf{Dataset construction.}
The database contains 52,010 records for AG News,
28,781 for Yahoo Answers, 100,000 for MedMCQA,
and 6,982 for CIFAR-100. Database records,
queries, and auxiliary information are generated
using disjoint pools. For the Zipf workloads,
records are sampled according to the specified
class proportions without replacement. The
natural query workload is sampled independently
of the database class distribution.

\noindent\textbf{Attack parameters.}
Table~\ref{tab:sage-parameters} summarizes the
default parameters. SAGE-Dist uses $d_{\mathrm{spec}}$
spectral dimensions and six KMeans candidates.
Its Sinkhorn refinement uses entropy parameter
$\varepsilon=0.05$, at most 30 refinement rounds,
and 200 Sinkhorn iterations per round.
SAGE-Known uses clamped label propagation with
class-mass normalization and the dataset-specific
parameters shown below. Propagation stops when
the relative score change is below $10^{-5}$
or after 500 iterations.

\begin{table}[ht]
\centering
\caption{Default attack parameters. $C$ is the
number of semantic categories.
$d_{\mathrm{spec}}$ is the SAGE-Dist
spectral embedding dimension.}
\label{tab:sage-parameters}
\scriptsize
\setlength{\tabcolsep}{3pt}
\renewcommand{\arraystretch}{1.08}
\begin{tabular}{lccccc}
\toprule
Dataset & $C$ & $d_{\mathrm{spec}}$
& $\alpha$ & $\gamma$ & $s$ \\
\midrule
AG News & 4 & $4C$ & 0.95 & 0.25 & 1.0 \\
Yahoo & 10 & $3C$ & 0.99 & 0.10 & 1.0 \\
MedMCQA & 21 & $3C$ & 0.99 & 0.10 & 1.0 \\
CIFAR-100 & 100 & $3C$ & 0.90 & 0.25 & 0.1 \\
\bottomrule
\end{tabular}
\end{table}

\subsection{Comparison Methods and Evaluation Protocol}
\label{app:attack-comparisons}

\noindent\textbf{Shared protocol.}
All methods use identical access patterns, database samples,
and query workloads. Known-label baselines receive the same
observed anchors $\Psi_V$ and Jaccard graph $W$ as SAGE-Known,
while prior-based baselines use the same estimated class
proportions as SAGE-Dist. Record-label accuracy is evaluated
over observed records, excluding disclosed labels.
Query categories are inferred using the majority vote in
Eq.~\eqref{eq:query-category-vote}.

\noindent\textbf{Prior-based baselines.}
Random assigns each observed record a uniformly sampled
label from $\mathcal C$. Prior-Sampling draws labels
independently from the estimated distribution
$\hat{\boldsymbol\pi}$. Prior-Majority assigns every
observed record the most probable class under
$\hat{\boldsymbol\pi}$. Ties are resolved by the
smallest class index.

\noindent\textbf{Nearest-known.}
Each unlabeled record is assigned the label of its
highest-weight adjacent anchor. Specifically, when
at least one anchor is adjacent to record $j$,
\begin{equation}
    \hat\lambda_j=\lambda_{j^*},\qquad
    j^*=\arg\max_{\substack{u\in K\cap V\\W_{ju}>0}}
    W_{ju}.
    \label{eq:app-nearest-known}
\end{equation}
Known anchors retain their original labels.
Records without adjacent anchors remain unassigned
until the shared completion step. Ties follow
the first maximum in sparse adjacency order.

\noindent\textbf{Louvain-known.}
We retain up to 20 highest-weight neighbors per node
and construct an undirected graph from their union.
We apply weighted Louvain community detection
with the default resolution parameter of $1$,
allowing the number of communities to be
determined automatically.
Each community containing observed
anchors receives their majority class, with ties
resolved by the smallest class index. Communities
without anchors remain unassigned. 
Unassigned records are completed using the
original Jaccard graph $W$.

\noindent\textbf{Standard graph propagation.}
Standard-LP applies iterative label propagation
using the random-walk transition matrix
$P_{\mathrm{rw}}=D^{-1}W$~\cite{zhu2002learning}.
Starting from the one-hot anchor matrix $Y$,
it repeatedly propagates class scores and clamps
anchor rows to their known labels. Zero-degree
rows remain zero. Iteration stops at a relative
tolerance of $10^{-6}$ or after 3,000 iterations.

Label-Spreading follows the propagation update
in Eq.~\eqref{eq:known-propagation} with
$\alpha=0.99$ and hard anchor clamping,
using a fixed-parameter variant
of~\cite{zhou2003learning}.
It stops at a relative tolerance of $10^{-5}$
or after 500 iterations.

\noindent\textbf{Completion and query voting.}
Unassigned records are completed through synchronous
weighted-neighbor voting, followed by the global
majority fallback described in
Section~\ref{sec:label-completion}.
If no labeled record is available, the fallback
selects class index $0$. Completed record labels
are then used for query-category majority voting.
The same completion procedure is applied to
SAGE and the corresponding baselines.

\subsection{Detailed Ablations and Fusion Sensitivity}
\label{app:sage-fusion}

\noindent\textbf{SAGE-Dist stages.}
Table~\ref{tab:dist-stage-ablation} reports the exact results
underlying Figure~\ref{fig:sage-ablation}(a). The three
variants use the same Jaccard co-retrieval graph, spectral
embedding dimension, and estimated class proportions.
A single candidate uses one KMeans candidate followed by
Hungarian size matching. Restart selection evaluates six
candidates and chooses the one with the smallest discrepancy
between named group sizes and the estimated proportions.
Full SAGE-Dist additionally applies Sinkhorn-based
record-level refinement. All predictions use the same
label-completion procedure.

\begin{table}[ht]
\centering
\caption{SAGE-Dist component ablations (record-label accuracy,
\%, mean $\pm$ standard deviation over ten runs).}
\label{tab:dist-stage-ablation}
\scriptsize
\setlength{\tabcolsep}{2.8pt}
\renewcommand{\arraystretch}{1.08}
\begin{tabular}{lrrrr}
\toprule
Variant & AG News & Yahoo & MedMCQA & CIFAR-100 \\
\midrule
Single candidate & $57.6\pm26.5$ & $32.1\pm6.7$ & $6.1\pm1.9$ & $2.0\pm0.8$ \\
+ Restart selection & $66.0\pm17.7$ & $34.2\pm4.3$ & $6.9\pm2.2$ & $1.8\pm0.4$ \\
+ Sinkhorn (full) & $66.9\pm18.0$ & $36.1\pm5.0$ & $7.3\pm2.3$ & $1.9\pm0.6$ \\
\bottomrule
\end{tabular}
\end{table}

\noindent\textbf{SAGE-Known fusion weight.}
Table~\ref{tab:known-fusion-sensitivity} varies the group-vote
weight $\gamma$ with 1\% randomly known record labels.
All variants use the same dataset-specific propagation
settings, Jaccard graph, and label-completion procedure.
For $\gamma>0$, the smoothing parameter is fixed to $s=1$
throughout this sweep. At $\gamma=0$, group votes are unused.
At $\gamma=1$, prediction relies on group votes, followed by
the same completion procedure. Known anchors retain their
supplied labels.

\begin{table}[ht]
\centering
\caption{SAGE-Known fusion-weight sensitivity at a 1\% known-label
budget (record-label accuracy, \%, mean $\pm$ standard deviation
over ten runs). Nonzero $\gamma$ uses $s=1$ in the sweep.
Full denotes the dataset-specific main configuration.}
\label{tab:known-fusion-sensitivity}
\scriptsize
\setlength{\tabcolsep}{3pt}
\renewcommand{\arraystretch}{1.06}
\begin{tabular}{rcccc}
\toprule
$\gamma$ & AG News & Yahoo & MedMCQA & CIFAR-100 \\
\midrule
$0$    & $82.0\pm0.9$ & $59.6\pm1.2$ & $45.3\pm0.8$ & $27.5\pm1.8$ \\
$0.10$ & $83.4\pm0.9$ & $\mathbf{64.9\pm1.4}$ & $\mathbf{47.0\pm0.7}$ & $27.5\pm1.8$ \\
$0.25$ & $\mathbf{84.3\pm0.6}$ & $63.3\pm1.8$ & $46.2\pm0.8$ & $27.6\pm1.8$ \\
$0.50$ & $78.6\pm2.0$ & $61.5\pm1.7$ & $43.6\pm0.8$ & $27.6\pm1.7$ \\
$0.75$ & $74.9\pm3.4$ & $60.6\pm1.8$ & $42.0\pm1.1$ & $\mathbf{27.8\pm1.7}$ \\
$1.00$ & $74.9\pm3.4$ & $59.8\pm1.3$ & $41.1\pm1.5$ & $25.7\pm2.6$ \\
\midrule
Full & $84.3\pm0.6$ & $64.9\pm1.4$ & $47.0\pm0.7$ & $27.8\pm1.7$ \\
\bottomrule
\end{tabular}
\end{table}

\subsection{Quiver Retrieval and Perturbation Details}
\label{app:quiver-protocol}

\noindent\textbf{Retrieval protocol.}
All four settings use the same sampled database,
query workload, and auxiliary knowledge as the
main evaluation. Exact uses inner-product top-$k$
search on normalized plaintext embeddings.
Quiver Plain performs plaintext HNSW retrieval
with \texttt{ef\_search}$=256$.
For Quiver DCPE, the client encrypts the database
and query embeddings before server-side HNSW
retrieval with the same search parameter.
The observed access patterns are obtained from
identifiers recorded in the Quiver server responses.
We check these traces against the identifiers
used by the attack. Recall@100 is the mean
fraction of retrieved identifiers also present
in the unperturbed exact top-100 results.

\noindent\textbf{Simulated perturbation.}
For normalized $x\in\mathbb R^d$, the local
simulation generates
\begin{equation}
    \widetilde{x}=s x+\eta,\qquad
    \eta=\frac{g}{\|g\|_2}
    \frac{s\beta}{4}\,u^{1/d},
    \label{eq:simulated-dcpe}
\end{equation}
where $g\sim\mathcal N(0,I_d)$,
$u\sim\operatorname{Unif}(0,1)$, and $s=1$.
Independent draws for database and query vectors
yield uniform perturbations within a $d$-dimensional
ball of radius $s\beta/4$. The perturbed vectors
are ranked by Euclidean distance using exact top-$k$
retrieval. This simulation is a retrieval-level
control, not a reproduction or security proof
of the deployed Quiver protocol.

\noindent\textbf{Perturbation sensitivity.}
Table~\ref{tab:quiver-beta} reports the full
$\beta$ sweep for simulated and Quiver DCPE retrieval.
Exact and Quiver Plain are unperturbed controls.
In dataset order (AG News, Yahoo Answers, MedMCQA,
and CIFAR-100), Quiver Plain attains Recall@100
of 99.3\%, 99.4\%, 99.4\%, and 100.0\%,
with SAGE-Dist accuracies of 72.5\%, 36.3\%, 6.5\%,
and 2.0\%, and SAGE-Known accuracies of
84.3\%, 65.0\%, 47.2\%, and 27.9\%.
Both attacks use their usual auxiliary-knowledge
settings. These comparisons are descriptive because
HNSW approximation and encryption-induced
perturbation may interact.

\begin{table}[t]
\centering
\caption{Retrieval Recall@100 and record-label accuracy
for the DCPE perturbation sweep (\%). Each entry is
the mean over the corresponding completed runs.
Sim. denotes local scale-and-perturb simulation,
and Quiver denotes encrypted Quiver HNSW retrieval.}
\label{tab:quiver-beta}
\scriptsize
\setlength{\tabcolsep}{4pt}
\renewcommand{\arraystretch}{1.06}
\begin{tabular}{llrrrrrr}
\toprule
& & \multicolumn{3}{c}{Simulated DCPE}
& \multicolumn{3}{c}{Quiver DCPE} \\
\cmidrule(lr){3-5}\cmidrule(lr){6-8}
Dataset & $\beta$ & Recall@100 & Dist & Known
& Recall@100 & Dist & Known \\
\midrule
AG News & 0.02 & 99.7 & 59.8 & 84.2 & 99.0 & 67.7 & 84.3 \\
 & 0.1 & 98.5 & 67.3 & 84.4 & 98.0 & 67.7 & 84.3 \\
 & 0.4 & 94.3 & 70.4 & 84.3 & 94.0 & 69.3 & 84.2 \\
 & 1.6 & 76.9 & 56.1 & 84.0 & 76.9 & 61.7 & 84.2 \\
\midrule
Yahoo & 0.02 & 99.6 & 34.9 & 64.8 & 99.1 & 35.0 & 65.0 \\
 & 0.1 & 98.4 & 32.5 & 64.7 & 98.0 & 32.7 & 65.0 \\
 & 0.4 & 93.8 & 34.4 & 64.7 & 93.6 & 32.6 & 64.8 \\
 & 1.6 & 74.8 & 34.7 & 63.9 & 74.8 & 32.8 & 64.7 \\
\midrule
MedMCQA & 0.02 & 99.7 & 7.1 & 47.1 & 99.1 & 5.5 & 47.1 \\
 & 0.1 & 98.5 & 6.5 & 47.1 & 98.1 & 6.6 & 47.2 \\
 & 0.4 & 94.4 & 5.1 & 47.2 & 94.1 & 5.8 & 47.1 \\
 & 1.6 & 76.8 & 5.4 & 46.5 & 76.9 & 5.4 & 46.9 \\
\midrule
CIFAR-100 & 0.02 & 99.6 & 1.9 & 27.8 & 99.5 & 2.1 & 27.9 \\
 & 0.1 & 98.1 & 2.2 & 27.8 & 98.1 & 1.7 & 27.9 \\
 & 0.4 & 92.9 & 2.0 & 27.9 & 92.9 & 1.5 & 27.7 \\
 & 1.6 & 68.8 & 1.8 & 27.1 & 68.8 & 1.9 & 27.1 \\
\bottomrule
\end{tabular}
\end{table}

\subsection{Identifier Rotation}
\label{app:rotation}

We evaluate idealized identifier rotation by
partitioning 2,000 queries into windows with
independent record pseudonyms. At a window
size of 50 queries, SAGE-Known achieves
76\% accuracy on AG News when known anchors
can be matched in every window, compared
with 29\% when anchors are restricted to the
first window. The corresponding results on
Yahoo Answers are 45\% and 19\%.
These findings suggest that preventing
cross-window record linkage can reduce
semantic inference, although other persistent
identifiers or metadata may still enable linkage.

\end{document}